\documentclass[12pt,letterpaper]{article}
\pdfoutput=1
\usepackage{fixltx2e}
\usepackage{textcomp}
\usepackage{fullpage}
\usepackage{amsfonts}
\usepackage{verbatim}
\usepackage[english]{babel}
\usepackage{pifont}
\usepackage{color}
\usepackage{setspace}
\usepackage{lscape}
\usepackage{indentfirst}
\usepackage[normalem]{ulem}
\usepackage{booktabs}
\usepackage[authoryear]{natbib}

\usepackage{float}
\usepackage{latexsym} 
\usepackage{url}
\usepackage{epsfig}
\usepackage{graphicx}
\usepackage{amssymb}
\usepackage{amsmath}
\usepackage{bm}
\usepackage{array}
\usepackage{mhchem}
\usepackage{ifthen}
\usepackage{caption}
\usepackage{amsthm}
\usepackage{amstext}

\usepackage{fullpage}
\usepackage{graphicx}
\usepackage{amsmath}
\usepackage{graphics} 
\usepackage{graphics}
\usepackage{pdfpages}
\usepackage{tikz}
\usepackage{subfig}      
\usetikzlibrary{positioning,shapes.geometric}

\usepackage{enumerate}

\newcommand{\R}{\ensuremath{\mathbb{R}}}
\newcommand{\std}[1]{\ensuremath{\mathbf{e}_{#1}}}

\def \R{\mathbb{R}}

\raggedright
\renewcommand{\section}[1]{%
\bigskip
\begin{center}
\begin{Large}
\normalfont\scshape #1
\medskip
\end{Large}
\end{center}}

\renewcommand{\subsection}[1]{%
\bigskip
\begin{center}
\begin{large}
\normalfont\itshape #1
\end{large}
\end{center}}

\renewcommand{\subsubsection}[1]{%
\vspace{2ex}
\noindent
\textit{#1.}---}

\renewcommand{\tableofcontents}{}

\bibpunct{(}{)}{;}{a}{}{,}  

\begin{document}
\begin{flushright}
Version dated: \today
\end{flushright}
\bigskip

\bigskip
\medskip
\begin{center}

\noindent{\Large \bf The effect of non-reversibility on inferring rooted phylogenies} 
\bigskip

\noindent {\normalsize \sc Svetlana Cherlin$^{1}$, Tom M.~W.~Nye$^2$, Sarah E.~Heaps$^{2}$, Richard J.~Boys$^2$,\\
Tom A.~Williams$^3$ and T.~Martin Embley$^4$}\\

\noindent {\small \it 
$^1$Institute of Genetic Medicine, Newcastle University, Newcastle upon Tyne, NE1 3BZ, U.K.\\
$^2$School of Mathematics \& Statistics, Newcastle University, Newcastle upon Tyne, NE1 7RU, U.K.\\ 
$^3$School of Earth Sciences, University of Bristol, Bristol, BS8 1RJ, U.K.\\
$^4$Institute for Cell and Molecular Biosciences, Newcastle University, Newcastle upon Tyne, NE2 4HH, U.K}.\\
\end{center}

\vspace{1in}

\subsubsection{Abstract}
Most phylogenetic models assume that the evolutionary process is stationary and reversible. As a result, the root of the tree cannot be inferred as part of the analysis because the likelihood of the data does not depend on the position of the root. Yet defining the root of a phylogenetic tree is a key component of phylogenetic inference because it provides a point of reference for polarising ancestor/descendant relationships and therefore interpreting the tree. In this paper we investigate the effect of relaxing the reversibility assumption and allowing the position of the root to be another unknown in the model. We propose two hierarchical models that are centred on a reversible model but perturbed to allow non-reversibility. The models differ in the degree of structure imposed on the perturbations. The analysis is performed in the Bayesian framework using Markov chain Monte Carlo methods.  We illustrate the performance of the two non-reversible  models in analyses of  simulated data sets using two types of topological priors. We then apply the models to a real biological data set, the radiation of polyploid yeasts, for which there is a robust biological opinion about the root position. Finally we apply the models to a second biological data set for which the rooted tree is controversial: the ribosomal tree of life. We compare the two non-reversible models and conclude that both are useful in inferring the position of the root from real biological data sets.
\\
\noindent (Keywords: rooting, phylogenetic tree, substitution model )\\

\vspace{1.5in}

\section{Introduction\label{sec:Intro}}

The root of a phylogenetic tree is fundamental to its biological interpretation, providing a critical reference point for polarising ancestor-descendant relationships and for determining the order in which key traits evolved along the tree \citep{embley_martin}. Despite its importance, most models of sequence evolution are based on homogeneous continuous time Markov processes (CTMPs) that are assumed to be stationary and time-reversible, with the mathematical consequence that the likelihood of a tree does not depend on where it is rooted. Therefore other methods are generally used to root evolutionary trees. The most common approach is to use an outgroup to the clade of interest, or ingroup; the root is then placed on the branch connecting the outgroup to the ingroup \citep{penny76, huelsenbeck2002}. However, this approach can be problematic if the outgroup is only distantly related to the ingroup because the long branch leading to the outgroup can induce phylogenetic artefacts such as long branch attraction (LBA), potentially interfering with the inference of ingroup relationships and the root position \citep{felsenstein1978, holland2003, bergsten2005}.  Indeed it has been proposed that the three-domains tree of life, in which Eukaryota represent the sister group to a monophyletic Archaea, could have resulted from LBA \citep{tourasse1999, williams2013}. Outgroup rooting is also difficult to apply to the question of rooting the universal tree, for which no obvious outgroup is available. One solution to this problem has been to use pairs of paralogous genes that diverged from each other before the last common ancestor of all cellular life, so that one paralogue can be used to root a tree of the other \citep{iwabe, brown_doolittle, hashimoto_hasegawa, baldauf}. However, for any given gene it is difficult to unambiguously establish that duplication took place before the divergence of the domains of life. The number of genes to which this technique can be applied is also limited. \par

\indent An alternative, but perhaps under-explored, approach to rooting trees is to take a model-based approach, adopting a substitution model in which changing the root position changes the likelihood of the tree. Focusing on homogeneous CTMPs, it is helpful to distinguish between the ideas of \textit{stationarity}, \textit{reversibility} and \textit{homogeneity}. We say that a model is \textit{homogeneous} if it can be characterised by a single instantaneous rate matrix that applies to the whole tree. A homogeneous model is termed \textit{reversible} if the rate matrix can be factorised into a symmetric matrix of exchangeability parameters and a diagonal matrix of stationary probabilities. Similarly we call a rate matrix \textit{reversible} if it permits such a factorisation. Finally a CTMP is \textit{stationary} if the probability of being in each state (e.g. each nucleotide for DNA) does not change over time and the probabilities of transitioning between states over some time interval depend only on the size of that interval and not on its position in time. It follows that all non-stationary models are also non-homogeneous, although the converse need not be true. Models in which one or more of these assumptions is relaxed can give rise to likelihood functions that depend on the position of the root.

For most models that allow root inference, the focus has been relaxing the assumption of homogeneity, typically assigning different reversible rate matrices to different parts of the tree. Generally, these models are non-stationary and allow variation in the theoretical stationary distribution across the tree. Some also allow variation in the exchangeability parameters \citep[][]{dutheilBoussau08} although, more commonly, they are fixed over all branches. For example, \citet{yangRoberts1995} assigned common exchangeabilities but a different composition vector to each edge of the tree. \citet{heaps2014} fitted a similar model in a Bayesian framework, but adopted a prior over composition vectors that allowed information to be shared between branches. Whilst biologically persuasive, such non-homogeneous models are, however, highly parameterised and efforts have been made to seek more parsimonious representations. \citet{yangRoberts1995} and \citet{foster2004} both considered models in which composition vectors are applied to groups of edges rather than to a single edge. \citet{blanquart2006} used a variation of this idea by assuming the compositional shifts occurred according to a Poisson process, independently of speciation events. In the context of nucleotide evolution, \citet{galtier1998} reduced the number of parameters in the model of \citet{yangRoberts1995} by using a model parameterised by a single G+C component, rather than three free parameters for the composition vector. But this inevitably came at the cost of a loss of information from the alignment. In a general setting that allowed different reversible or non-reversible rate matrices to be assigned to each edge of the tree, \citet{jayaswal2011} devised a heuristic to reduce the number of rate matrices using the distances between them as a similarity criteria, and forcing the most similar rate matrices to be identical. However, given the speculative nature of the model search, the algorithm offered no assurance of identifying a global optimum.

In spite of these moves towards parsimony, non-homogeneous models remain substantially more highly parameterised than their homogeneous counterparts. This makes model-fitting computationally challenging, often limiting inference to fixed unrooted trees \citep[e.g.][]{dutheilBoussau08,jayaswal2011} or alignments on a small number of taxa \citep[e.g.][]{heaps2014}. In this paper we take a Bayesian approach to inference and focus on rooting using a \textit{homogeneous} and stationary, but non-reversible, model that requires only \textit{one} rate matrix. This model has previously been explored by \citet{huelsenbeck2002}, however we build on the work in a number of ways. First, we do not fix the unrooted topology and extend the inferential algorithm to allow inference of rooted trees. This allows us to present a more complete summary of the posterior over root positions and to demonstrate the sensitivity of the analysis to different topological priors. Additionally, whilst \citet{huelsenbeck2002} only considered small alignments of up to nine taxa, we consider more compelling analyses with data sets of up to 36 taxa. Finally, \citet{huelsenbeck2002} used a so-called non-informative prior on the rate matrix, with independent uniform distributions for each off-diagonal element. We incorporate prior structure and consider two hierarchical priors that are centred on a standard reversible rate matrix but allow non-reversible perturbations of the individual elements. Our two priors differ in the structure of the perturbation. We test our hierarchical models on simulated data and on a real biological data set for which there is a robust biological opinion about the position of the root. Finally, we apply the models to an open question in biology: the root of the tree of life.

\section{New approaches}

\subsection{Top level model description}
We consider a number of aligned homologous sequences and aim to infer the evolutionary relationships among these sequences. These relationships can be described in the form of a bifurcating tree, where  each edge represents the period of time over which point mutations accumulate, and each bifurcation represents a speciation event. The nucleotides at each site of a sequence alignment on $n$ taxa can be thought of as independent realisations of a random variable $X = (x_1, ..., x_n)^T$ on a discrete space where $x_i \in \Omega$ and $\Omega = \{$A, G, C, T$\}$, for $i=1,\ldots,n$. The evolutionary process operating along each edge of the tree is described by a homogeneous CTMP, where the future value of a nucleotide at any given site depends on its current value only and does not depend on its past values given this current value, that is 
\begin{align*}
\Pr (& X(t) =j|X(t_1)=i_1,~X(t_2)=i_2, ~\dots,~X(t_n)=i_n)\\
= & \Pr(X(t)=j|X(t_n)=i_n), 
\end{align*}
where $t > t_n > t_{n-1} > ... > t_2 > t_1$. The process can therefore be specified by a transition matrix $P(\ell) =  \{p_{ij} (\ell)\}$ whose elements $p_{ij}(\ell)$ represent the probabilities of changing from one nucleotide to another over a branch of length $\ell$. Equivalently we can represent the process through an instantaneous rate matrix $Q$,  where $P(\ell) =  \exp(Q\ell)$. The off-diagonal elements of $Q$ represent an instantaneous rate of change from one nucleotide to another during an infinitesimal period of time.
The diagonal elements are specified so that every row sums to zero.
If branch lengths need to be expressed in terms of expected number of substitutions per site then the $Q$ matrix has to be rescaled so that $-\sum Q_{ii}\pi_{Q,i} = 1$, where $\boldsymbol\pi_Q = (\pi_{Q,A}, \pi_{Q,G}, \pi_{Q,C}, \pi_{Q,T})$ is the theoretical stationary distribution of the process, which can be calculated from $Q$.

Most phylogenetic models are time-reversible. Reversibility implies that
\begin{equation*}
\pi_{Q,i}p_{ij} = \pi_{Q,j}p_{ji}
\end{equation*}
and allows the rate matrix\label{pg:2.6ii} to be represented in the form $Q = S \Pi$, where $S$ is a symmetric matrix containing the exchangeability parameters $\rho_{ij}$, $i \ne j$, as the off-diagonal elements with $\rho_{ij} = \rho_{ji}$,  and  $ \Pi = \text{diag}(\boldsymbol\pi_Q)$ is a diagonal matrix containing the elements of $\boldsymbol\pi_Q$.
While the reversibility assumption makes statistical models simpler, it has no biological justification, and is applied for computational convenience only. Indeed, there is often evidence of non-reversibility in biological data sets \citep{squartini2008,sumner2015}.
 
The most common reversible rate matrix, with six exchangeability parameters, is the general time-reversible (GTR) model \citep{tavare86}. The HKY85 model \citep{hasegawa85} is a widely used special case with only two distinct $\rho_{ij}$, one of which is fixed to prevent arbitrary rescaling of the $Q$ matrix. The rate matrix $Q$ of this model is then specified by the compositional frequency vector $\boldsymbol\pi = (\pi_A, \pi_G, \pi_C, \pi_T)$ and by the transition-transversion rate ratio $\kappa$ as 
\begin{equation*}
Q = \begin{pmatrix}
\star &    \kappa\pi_G & \pi_C & \pi_T \\
\kappa\pi_A &\star & \pi_C & \pi_T \\
\pi_A & \pi_G & \star & \kappa\pi_T \\
\pi_A & \pi_G & \kappa\pi_C &\star \end{pmatrix}.
\end{equation*}
Here the symbol $\star$ is used to indicate that the diagonal elements are specified such that every row sums to zero.

We consider two Bayesian hierarchical models that are both non-reversible and therefore based on an unstructured rate matrix $Q$. The models differ in the prior they assign to its off-diagonal elements $q_{ij}$. In each case the prior treats each $q_{ij}$ as a log-normal perturbation of the corresponding element of the unknown rate matrix of a HKY85 model. The first hierarchical model, henceforth called the NR model, utilises one perturbation component, while the more complex model, henceforth called the NR2 model, utilises two perturbation components. The variances of the perturbations are unknown and can provide a measure of the evidence of non-reversibility in the data.

In both models we assume that the variation between the overall rate of substitution events at sites can be modelled by a Gamma distribution with mean equal to 1 \citep{yang1993}. For computational convenience we approximate the continuous $\mathrm{Ga}(\alpha, \alpha)$ distribution with a discrete $\mathrm{Ga} (\alpha, \alpha)$ distribution with four categories \citep{yang1994}.

\subsection{Top level prior distribution}
\subsubsection{NR model}

We denote the off-diagonal elements of the rate matrix of the NR model by $q_{ij}$, and the off-diagonal elements of the rate matrix of the HKY85 model by $q_{ij}^H$, $i \neq j$, so for instance $q_{12}^H=\kappa\pi_G$.
The non-reversibility of the NR model is achieved by a log-normal perturbation of the off-diagonal elements of the rate matrix $Q^H$ using a perturbation component $\sigma$ as represented in the following directed acyclic graph (DAG):

\begin{center}
\begin{tikzpicture}[%
->,shorten >=2pt,>=stealth,node distance=0.5cm,pil/.style={->,thick,shorten =2pt,}]

\node[circle,draw,minimum size=1.2cm] (2) {$Q$} ;
\node[circle,draw,minimum size=1.2cm] [below=of 2] (3) {$\sigma$} ;
\node[circle,draw,double,double distance=1pt,minimum size=1.2cm] [left =of 2] (1) {$Q^H$} ;
\node[circle,draw,minimum size=1.2cm] [right=of 2] (4) {Data};
\node[circle,draw,minimum size=1.2cm] [below=of 4] (5) {$\tau, \boldsymbol\ell$};
\node[circle,draw,minimum size=1.2cm] [right=of 4] (6) {$\alpha$}  ;
\node[circle,draw,minimum size=1.2cm] [below=of 1] (7) {$\boldsymbol\pi,\kappa$};

\draw[->] (1.east) --(2.west);
\draw[->] (2.east) -- (4.west);
\draw[->] (7.north) --(1.south);
\draw[->] (3.north) --(2.south);
\draw[->] (5.north) --(4.south);
\draw[->] (6.west) --(4.east);

\end{tikzpicture}
\end{center}

DAGs are a useful way of representing (especially hierarchical) models graphically. In a DAG, the nodes represent random variables and the directed arrows are used to indicate the order of conditioning when factorising the joint probability density of all the nodes.  A double circle around a node indicates deterministic dependence; in this case $Q^H$ is completely determined once $\boldsymbol\pi$ and $\kappa$ are known. In the DAG above, $\alpha$ is the across-site heterogeneity parameter, $\tau$ is the rooted topology and $\boldsymbol\ell$ are the branch lengths.

Working element-wise on a log-scale, the off-diagonal elements of the rate matrix of the NR model can be expressed as, for $i \neq j$
\begin{equation*} 
\log q_{ij} =  \log q^{H}_{ij} + \epsilon_{ij},  
\end{equation*}
where the $\epsilon_{ij}$ are independent $N(0,\sigma^2)$ quantities. Here the perturbation standard deviation $\sigma$ represents the extent to which $Q$ departs from a HKY85 structure: the larger its value, the greater the degree of departure. This parameter is treated as an unknown quantity whose value we learn about during the analysis. The unknowns of the hierarchical model therefore comprise: the composition vector $\boldsymbol\pi$, the transition-transversion rate ratio $\kappa$, the perturbation standard deviation $\sigma$, the off-diagonal elements of the rate matrix $Q$, the shape parameter $\alpha$, the branch lengths $\boldsymbol\ell$ and the rooted topology $\tau$. We express our initial uncertainty about these unknown parameters though a prior distribution that takes the form
\begin{equation}\label{eq:priorNR}
\pi (\boldsymbol\pi, \kappa, \sigma, Q, \alpha, \boldsymbol \ell, \tau) = 
\pi(Q|\boldsymbol\pi, \kappa, \sigma)\pi(\boldsymbol\pi, \kappa, \sigma, \alpha, \boldsymbol \ell, \tau)
\end{equation}
in which the top-level prior density $\pi(Q|\boldsymbol\pi, \kappa, \sigma)$ has been described above. The bottom level density $\pi(\boldsymbol\pi, \kappa, \sigma, \alpha, \boldsymbol \ell, \tau)$ will be described in the subsection \textit{Bottom level prior distribution}.

\subsubsection{NR2 model}

Under the NR model, departures from HKY85 structure could lead to a non-reversible model or simply a general time-reversible rate matrix. \label{pg:2.6iii} As such the two types of deviation are confounded and so for any given data set, learning that $\sigma$ is large  does not necessarily provide evidence of non-reversibility. The NR2 model addresses this issue, thereby aiding model interpretation, by using a two-stage process to perturb the underlying HKY85 rate matrix $Q^H$. 
The first perturbation is within the space of GTR matrices, perpendicular to the subspace of HKY85 matrices, leading to a reversible rate matrix denoted $Q^R$. 
The second perturbation acts on $Q^R$ and is within the space of general rate matrices but perpendicular to the subspace of GTR matrices, leading to a general non-reversible rate matrix denoted $Q$.
These two random perturbations have different variance parameters $\sigma_R^2$ and $\sigma_N^2$ respectively. Biologically\label{pg:2.3ii}, the variance parameter $\sigma_R^2$ represents the extent to which the data contradict the assumption of a common rate of transition and a common rate of transversion. Similarly, the variance parameter $\sigma_N^2$ provides a measure of the evidence in the data for the directionality of time.

The general structure of this model can be represented by the following DAG:

\begin{center}
\begin{tikzpicture}[%
->,shorten >=2pt,>=stealth,node distance=0.5cm,pil/.style={->,thick,shorten =2pt,}]

\node[circle,draw,minimum size=1.2cm] (2) {$Q^R$} ;
\node[circle,draw,minimum size=1.2cm] [below=of 2] (5) {$\sigma_R$} ;
\node[circle,draw,double,double distance=1pt,minimum size=1.2cm] [left =of 2] (1) {$Q^H$} ;
\node[circle,draw,minimum size=1.2cm] [right=of 2] (3) {$Q$}  ;
\node[circle,draw,minimum size=1.2cm] [below=of 3] (6) {$\sigma_N$};
\node[circle,draw,minimum size=1.2cm] [right=of 3] (4) {Data};
\node[circle,draw,minimum size=1.2cm] [below=of 4] (7) {$\tau, \boldsymbol\ell$};
\node[circle,draw,minimum size=1.2cm] [right=of 4] (9) {$\alpha$}  ;
\node[circle,draw,minimum size=1.2cm] [below=of 1] (8) {$\boldsymbol\pi,\kappa$};

\draw[->] (1.east) -- (2.west);
\draw[->] (2.east) -- (3.west);
\draw[->] (3.east) -- (4.west);
\draw[->] (5.north) --(2.south);
\draw[->] (6.north) --(3.south);
\draw[->] (7.north) --(4.south);
\draw[->] (8.north) --(1.south);
\draw[->] (9.west) --(4.east);

\end{tikzpicture}
\end{center}
	
\noindent
The two-stage perturbation procedure is explained further in Appendix 1. The unknown parameters in the NR2 model are therefore: the composition vector $\boldsymbol\pi$, the transition-transversion rate ratio $\kappa$, the perturbation standard deviation on the reversible plane $\sigma_R$, the perturbation standard deviation on the non-reversible plane $\sigma_N$, the shape parameter $\alpha$, the branch lengths $\boldsymbol\ell$ and the rooted topology $\tau$. We also have latent variables comprising $\nu_1,\dots,\nu_5$  for the reversible perturbation,  and $\eta_1,~\eta_2,~\eta_3$ for the non-reversible perturbation (see Appendix 1). The prior distribution of these unknowns takes the form
\begin{equation} \label{eq:priorNR2}
\begin{split}
\pi (&\boldsymbol\pi, \kappa,  \sigma_R, \sigma_N, \boldsymbol\nu, \boldsymbol\eta, \alpha, \boldsymbol \ell, \tau) \\ 
&= \pi(\boldsymbol\nu|\sigma_R) \pi(\boldsymbol\eta|\sigma_N) \pi(\boldsymbol\pi, \kappa, \sigma_R, \sigma_N, \alpha, \boldsymbol\ell, \tau). 
\end{split}
\end{equation}  
where the top-level prior distributions with densities $\pi(\boldsymbol\nu|\sigma_R)$ and $\pi(\boldsymbol\eta|\sigma_N)$ are $\nu_i \sim N(0, \sigma^2_R)$ for $i=1,\dots,5$ independently,  and $\eta_i \sim N(0, \sigma^2_N)$ for $i=1,~2,~3$ independently (see Appendix 1). The bottom level density $\pi(\boldsymbol\pi, \kappa, \sigma_R, \sigma_N, \alpha, \boldsymbol \ell, \tau)$ will be described in the following subsection.

\subsection{Bottom level prior distribution}
\subsubsection{NR model}
The bottom-level prior density $\pi(\boldsymbol\pi, \kappa, \sigma, \alpha, \boldsymbol \ell, \tau)$ from~\eqref{eq:priorNR} takes the form
\begin{equation*}
\pi(\boldsymbol\pi, \kappa, \sigma, \alpha, \boldsymbol \ell, \tau) = 
\pi(\boldsymbol\pi) \pi(\kappa) \pi(\sigma) \pi(\alpha)\pi(\boldsymbol\ell)\pi(\tau)
\end{equation*}
to reflect our initial assessment of independence between these parameter blocks. 

The composition vector $\boldsymbol\pi$ is defined on the four-dimensional simplex, that is, it has four positive elements, constrained to sum to one. We choose to assign it a Dirichlet prior, $\boldsymbol\pi \sim \mathcal{D} (a_{\pi}\boldsymbol\pi_0)$, where $\boldsymbol\pi_0 = (0.25, 0.25, 0.25, 0.25)$ is the mean and $a_{\pi}$ is a concentration parameter (we take $a_{\pi} = 4$). This prior is exchangeable with respect to the nucleotide labels.
We adopt a log-normal prior for the transition-transversion rate ratio
$\kappa \sim LN(\log\kappa_0, \nu^2)$, where $\kappa_0$ = 1 and $\nu$ = 0.8.
The parameters of the prior for  $\kappa$ represent our belief that the probability of $\kappa$ exceeding 2 is 0.2, i.e  $\Pr(\kappa < 2) = 0.8$. The perturbation parameter $\sigma$ is assigned an Exponential prior
$\sigma \sim \mathrm{Exp}(\gamma)$, where the rate $\gamma = 2.3$ reflects our prior belief that the probability of $\sigma$ exceeding 1 is 0.1, i.e $\Pr(\sigma < 1)  = 0.9$. Together with the rest of our hierarchical specification, this choice induces a prior for the stationary distribution $\boldsymbol\pi_Q$ in which little density is assigned to vectors where some characters are heavily favoured over the others.

The branch lengths are assigned independent Exponential priors
$\ell_i \sim \mathrm{Exp}(\mu)$, where $i = 1, ..., k$ and $k$ is the number of edges. The rate $\mu$ equals 10, so that $\mathrm{E}(\ell_i)  = 0.1$. 
The shape parameter $\alpha$ is assigned a Gamma prior, 
$\alpha \sim \mathrm{Ga}(10, 10)$, which ensures the expected substitution rate in the $\mathrm{Ga}(\alpha, \alpha)$ model for site-specific substitution rates is modestly concentrated around 1.

We define a \emph{root type} as the number of species on each side of the root. For example, the root type $1 : (n-1)$ represents a root split on a pendant edge, $2 : (n-2)$ represents a root split between two taxa and all others, etc. A uniform prior over rooted topologies assigns a prior probability of more than 0.5 to root splits of the type $1 : (n-1)$, in other words, to roots on pendant edges. We felt that deeper roots are generally more biologically plausible and should be assigned higher prior mass, whilst still retaining a diffuse initial distribution. We therefore chose to assign the rooted topology a prior according to the Yule model of speciation, which assumes that at any given time each of the species is equally likely to undergo a speciation event. This generates a biologically defensible prior in which all root types receive the same prior probability if $n$ is odd, and a near uniform distribution if $n$ is even, but with $n/2 : n/2$ root types receiving half the prior probability of the other root types. The probability of generating a $n$-species tree $T$ under the Yule distribution is calculated by dividing the number of labelled histories for the tree $T$ by the total number of all possible labelled histories on $n$ species  \citep{steel}. This probability depends on the complete rooted topology and therefore has to be re-calculated at every iteration of the Metropolis Hastings algorithm used for inference. To save computational time, we therefore additionally introduce an approximation to the Yule prior, which we term the \emph{structured uniform prior}, that assigns equal prior probability to all root types. In order to sample a rooted topology from this distribution we first sample a root type uniformly. We then sample uniformly from the set of all rooted topologies with that root type. Computationally, this prior is more convenient than the Yule prior because its mass function is independent of the particular unrooted topology and only considers the root split. It also has the advantage of being uniform on root types for all $n$. Posterior sensitivity to the choice of topological prior will be discussed in the \textit{Analysis of experimental data} subsection.

\subsubsection{NR2 model}

The bottom-level prior density \newline
$\pi(\boldsymbol\pi, \kappa, \sigma_R, \sigma_N, \alpha, \boldsymbol \ell, \tau)$ from \eqref{eq:priorNR2} takes the form
\begin{align*}
\pi(& \boldsymbol\pi, \kappa, \sigma_R, \sigma_N, \alpha, \boldsymbol\ell, \tau) \\&= \pi(\boldsymbol\pi) \pi(\kappa) \pi(\sigma_R) \pi(\sigma_N) \pi(\alpha) \pi(\boldsymbol\ell) \pi(\tau).
\end{align*}

The rate heterogeneity parameter $\alpha$, branch lengths $\boldsymbol\ell$, rooted topology $\tau$ and the parameters $\boldsymbol\pi$ and $\kappa$ of the reversible $Q^H$  matrix are assigned the same priors as those used for the NR model. Both perturbation standard deviations are assigned the same prior as their analogue, $\sigma$, in the NR model, i.e. $\sigma_R \sim \mathrm{Exp}(2.3)$ and $\sigma_N \sim \mathrm{Exp}(2.3)$.

\section{Results}
Taking a Bayesian approach to inference, we fitted the NR and NR2 models to the data sets described in this section using a Markov chain Monte Carlo algorithm. Full details of the inferential procedure are provided in the \textit{Materials and methods} section.

\subsection{Analysis of simulated data}

Our simulations aim to explore two aspects: (i) the effect of different levels of non-reversibility in the data on root inference; (ii) the effect of different topologies and branch lengths on root inference.

\subsubsection{Different levels of non-reversibility in the data}

Here we explore the posterior when the NR and NR2 models are fitted to simulated data that contain different levels of non-reversibility. The tree used to simulate the data is a random 30-taxon tree (generated under the Yule birth process), with the branch lengths simulated from Ga(2,20). The lengths of the branches adjacent to the root are simulated from Ga(1,20) such that the combined length of these two branches corresponds to a Ga(2,20) random variable (Supplementary Fig. 1).
In order to simulate the alignments, we first fix the underlying reversible HKY85 rate matrix ($Q^H$ matrix) using the  values $\boldsymbol\pi = (0.25, 0.25, 0.25, 0.25)$ and $\kappa =2$. We then apply different types of perturbation to the $Q^H$ matrix.

\paragraph{NR model.}

Five different values of the perturbation standard deviation $\sigma$ were used to simulate the data: $\sigma$ = 0, 0.05, 0.1, 0.2, 0.3. For each value of $\sigma$ nine different data sets of length 2000 bp were simulated, the first five having different rate matrices (data sets 1 - 5), and the last five having the same rate matrix (data sets 5 - 9). Thus the former five data sets have different stationary distributions $\boldsymbol\pi_Q$,  while the latter five data sets have the same stationary distribution. This type of alignment simulation allows us to investigate different sources of variability in the data. All the alignments were simulated using a gamma shape heterogeneity parameter generated from $\mathrm{Ga}(10, 10).$ Note that the case of $\sigma=0$ corresponds to the reversible HKY85 model. The other values of $\sigma$ were chosen so that the prior for the stationary distribution induced  by the log-normal perturbation would be in the range of values estimated for real data; as $\sigma$ increases, significant support is given to highly biased compositions, and for $\sigma > 0.3$ these are biologically unrealistic (Supplementary Fig. 2). 

To provide a consistent measure of non-reversibility across both the NR and NR2 models, we consider the value of Huelsenbeck's $I$ statistic ($I = \sum_{ij}|\pi_iq_{ij}-\pi_jq_{ji}|$,  \citet{huelsenbeck2002}). Under a reversible model, $\pi_iq_{ij} = \pi_jq_{ji}$ for all $i \ne j$, and so $I=0$. However, $I$ is strictly positive for non-reversible models, with larger values indicating a greater degree of non-reversibility. The values of Huelsenbeck's $I$ statistic for the models used to generate the data in these experiments are shown in Table~\ref{tab:istatNR}.

\begin{table*}[!t]
\begin{center}
\caption{Values of Huelsenbeck's $I$ statistic for the $Q$ matrices used in the simulations with the NR model. By design, there is a strong positive correlation between $\sigma$ and $I$.}
\label{tab:istatNR}
\begin{tabular}{ c c c c c c   } \hline                                   
Data Set             & $\sigma$ = 0 & $\sigma$ = 0.05 & $\sigma$ = 0.1 & $\sigma$ = 0.2 & $\sigma$ = 0.3 \\ \hline
1  &0.0000  &0.0398  &0.0534  &0.1281  &0.3191\\
2  &0.0000  &0.0287  &0.1428  &0.2553  &0.1483\\
3  &0.0000  &0.0407  &0.0438  &0.1105  &0.0991\\
4  &0.0000  &0.0500  &0.1163  &0.1292  &0.1383\\
5-9  &0.0000  &0.0628  &0.0322  &0.0827  &0.1088\\ \hline                  
\end{tabular}
\end{center}
\end{table*}

Table \ref{tab:postNR} summarises the marginal posterior probabilities of the correct root split and the posterior means for Huelsenbeck's $I$ statistic for the data simulated with the NR model and analysed under the Yule prior (the posterior distributions of the root splits are shown in Supplementary Fig. 3). When $\sigma=0$ the posterior of the root splits is identical to the prior (not shown) because the data contain no information about the root. As $\sigma$ increases, the root is often inferred better, with $\sigma=0.3$ demonstrating the best root inference of all analysed values of $\sigma$. However, the analyses of nine simulated data sets for each value of $\sigma$ do not show identical behaviour. There is substantial variability between the data sets, even those simulated with the same rate matrix, and the true root split is not inferred well in all experiments. The true unrooted topology, however, is inferred with posterior probability close to one in all cases (Supplementary Fig. 4). This suggests that in addition to inferring the unrooted topology, we can also use the NR model to extract some information about the root. Moreover, as expected, the greater the degree of non-reversibility, the stronger the signal from the data.

\begin{table*}[!bt]
\begin{center}
\caption{Marginal posterior probabilities of the correct root split for the simulations with the NR model and the Yule prior. The posterior means for Huelsenbeck's $I$ statistic are indicated in parentheses. When the correct root split is a modal root split, the corresponding marginal posterior probability appears in bold.}
\label{tab:postNR}
\begin{tabular}{ c c c c c c   } \hline                                   
Data Set             & $\sigma$ = 0 & $\sigma$ = 0.05 & $\sigma$ = 0.1 & $\sigma$ = 0.2 & $\sigma$ = 0.3 \\ \hline
1 & 0.08 (0.02) & 0.09 (0.02)  & 0.10 (0.07)  & {\bf{0.40}} (0.16) & {\bf{0.88}} (0.30) \\ 
2 & 0.09 (0.04) & 0.12 (0.03)  & 0.19 (0.13)  & 0.02 (0.26) & {\bf{0.44}} (0.19)    \\ 
3 & 0.10 (0.03) & 0.10 (0.03) & 0.07 (0.04) & 0.03 (0.11) & 0.22 (0.12) \\ 
4 & 0.07 (0.02) & 0.07 (0.07)  & 0.17 (0.12) & {\bf{0.23}} (0.08) & 0.16 (0.15) \\
5 & 0.08 (0.04)  & 0.17 (0.05) & 0.04 (0.05) & 0.09 (0.08) & {\bf{0.32}} (0.10)\\ 
6 & 0.08 (0.02)  & {\bf{0.21}} (0.05) & 0.06 (0.02) & 0.13 (0.08) & {\bf{0.50}} (0.14)\\ 
7 & 0.09 (0.02)  & {\bf{0.21}} (0.06) & 0.08 (0.01) & {\bf{0.21}} (0.09) & 0.10  (0.11)\\ 
8 & 0.08 (0.04)  & 0.23 (0.07) & 0.06 (0.03) & 0.16 (0.10) & 0.03 (0.08)\\ 
9 & 0.10 (0.03)  & 0.23 (0.04) & 0.11 (0.04) & 0.16 (0.08) & 0.14 (0.10)\\ \hline                  
\end{tabular}
\end{center}
\end{table*}

In order to evaluate the sensitivity of the analysis to the topological prior, the same analysis was performed using the structured uniform prior (Supplementary Tab. 1, Supplementary Fig. 5 and 6). This analysis gave very similar results, as we might expect given the similarity between the two priors.

\paragraph{NR2 model.}

The simulations were performed in a similar manner as for the NR model. Nine alignments were created for each of five values of $\sigma_N$ = 0, 0.1, 0.25, 0.5, 1.0. In all the simulations we used the same value for the reversible perturbation, $\sigma_R = 0.1$. Note that the case of $\sigma_N = 0$ corresponds to the GTR model. The values of $\sigma_N$ = 0.1, 0.25, 0.5, 1.0 were chosen so that in the prior for the stationary distribution, some nucleotides are not heavily favoured over the others (Supplementary Fig. 7). We note that this type of perturbation allows us to use larger values of $\sigma_N$ in comparison to the values of $\sigma$ in the NR model, while still maintaining a realistic stationary distribution. This, in turn, means we can simulate data from models with a greater degree of non-reversibility and, correspondingly, larger values of Huelsenbeck's $I$ statistic. This is illustrated in Table~\ref{tab:istatNR2} which displays the values of Huelsenbeck's $I$ statistic for the models used to generate the data in these experiments.  As for the NR model, for each value of $\sigma_N$ the first five alignments were simulated from different rate matrices (data sets 1 - 5), while the last five alignments were simulated from the same rate matrix (data sets 5 - 9). All the alignments were simulated using a gamma shape heterogeneity parameter simulated from $\mathrm{Ga}(10, 10).$

\begin{table*}[!t]
\begin{center}
\caption{Values of Huelsenbeck's $I$ statistic for the $Q$ matrices used in the simulations with the NR2 model. By design, there is a strong positive correlation between $\sigma_N$ and $I$.}
\label{tab:istatNR2}
\begin{tabular}{ c c c c c c   } \hline                                   
Data Set             & $\sigma_N$ = 0 & $\sigma_N$ = 0.1 & $\sigma_N$ = 0.25 & $\sigma_N$ = 0.5 & $\sigma_N$ = 1.0 \\ \hline
1 &0.0000 &0.0550 &0.2327 &0.3282 &1.0416\\
2 &0.0000 &0.0366 &0.1871 &0.4423 &0.9019\\
3 &0.0000 &0.0737 &0.3297 &0.4699 &0.7494\\
4 &0.0000 &0.0538 &0.1675 &0.3654 &0.7282\\
5-9 &0.0000 &0.1012 &0.3541 &0.4402 &0.9948\\ \hline                 
\end{tabular}
\end{center}
\end{table*}

Table \ref{tab:postNR2} summarises the marginal posterior probabilities of the correct root split and the posterior means for Huelsenbeck's $I$ statistic for the data simulated with the NR2 model and analysed under the Yule prior (the posterior distributions of the root splits are shown in Supplementary Fig. 8). As with the NR model, when  $\sigma_N=0$ the posterior probability of the root splits is very similar to the prior (not shown). This is because the data contain no information about the root position when simulated under a reversible model. As $\sigma_N$ increases, the root is inferred better, with $\sigma_N=1$ demonstrating the best root inference of all the values of $\sigma_N$ analysed. For the simulations under the NR2 model, the posteriors are more concentrated around the true root position than they had been for the simulations under the NR model. However, comparing the values for Huelsenbeck's $I$ statistic in Tables~\ref{tab:istatNR} and \ref{tab:istatNR2}, this is simply because the data simulated under the NR2 model generally had a higher degree of non-reversibility. Indeed, when fitting the NR model to the data simulated under the NR2 model, we obtained very similar root inferences to those summarised in Table~\ref{tab:postNR2}, with strong posterior support for the correct root position for large $\sigma_N$. 

\begin{table*}[!bt]
\begin{center}
\caption{Marginal posterior probabilities of the correct root split for the simulations with the NR2 model and the Yule prior. The posterior means for Huelsenbeck's $I$ statistic are indicated in parentheses. When the correct root split is a modal root split, the corresponding marginal posterior probability appears in bold.}
\label{tab:postNR2}
\begin{tabular}{ c c c c c c   } \hline                                   
Data Set             & $\sigma_N$ = 0 & $\sigma_N$ = 0.1 & $\sigma_N$ = 0.25 & $\sigma_N$ = 0.5 & $\sigma_N$ = 1.0 \\ \hline
1 & 0.07 (0.24) & 0.11 (0.04) & {\bf{0.63}} (0.23) & {\bf{0.92}} (0.35)  & {\bf{0.99}} (1.03)\\ 
2 & 0.09 (0.25) & 0.08 (0.03) & 0.07 (0.16) &{\bf{0.87}} (0.41) & {\bf{0.95}} (0.76)\\
3 & 0.05 (0.05) & 0.13 (0.08) & 0.20 (0.36) & 0.29 (0.49) & {\bf{0.98}} (0.69)\\
4 & 0.06 (0.07) & 0.07 (0.03) & 0.09 (0.21) & {\bf{0.63}} (0.33) & {\bf{0.99}} (0.77)\\
5 & 0.08 (0.02) & 0.22 (0.10) & {\bf{0.34}} (0.34) & {\bf{0.91}} (0.51) & {\bf{1.00}} (1.03)\\
6 & 0.07 (0.02) & 0.13 (0.04) & 0.21 (0.37) &{\bf{0.92}} (0.51) & {\bf{0.99}} (1.00)\\
7 & 0.07 (0.02) & 0.03 (0.13) & {\bf{0.48}} (0.32) &{\bf{0.88}} (0.46)  & {\bf{0.95}} (0.94)\\
8 & 0.08 (0.03) & 0.18 (0.08) & {\bf{0.36}} (0.36) & {\bf{0.97}} (0.45) & {\bf{0.99}} (1.02)\\
9 & 0.08 (0.02) & 0.09 (0.07) & 0.23 (0.32)& {\bf{0.65}} (0.44) & {\bf{0.99}} (0.98)\\ \hline                  
\end{tabular}
\end{center}
\end{table*}

In terms of inference for the unrooted tree, the true topology had posterior probability close to 1 in all cases (Supplementary Fig. 9). The analysis of the same data sets performed with the structured uniform prior showed similar results (Supplementary Tab. 2, Supplementary Fig. 10 and 11).

\subsubsection{Different topologies and branch lengths}

In a Bayesian analysis, the posterior distribution reflects information from both the prior and the data. When the prior and likelihood are comparably concentrated, but in conflict, the posterior can only represent a middle ground. In phylogenetics, inferences can be highly sensitive to the choice of prior for branch lengths and the topology itself \citep[][]{yangRannala2005,alfaroHolder2006}.

Motivated by the kinds of conflicts that are likely to arise in the analysis of real biological data, we consider the robustness of posterior root inferences to conflicting prior and likelihood information concerning the rooted topology and branch lengths. In our analyses we adopt the commonly used Exp(10) prior for branch lengths and a Yule prior (or the approximating stuctured uniform prior) over rooted topologies. An Exp(10) prior for branch lengths asserts a strong prior belief that edges will be reasonably short. Therefore, given an unrooted topology that contains a long branch, the prior will typically support placement of the root midway along this branch in order to break it up into two shorter ones. The Yule prior for rooted topologies assigns a (near) uniform distribution to all root types. However, there are generally many more trees of unbalanced types, like $1:n-1$, than there are of more balanced types like $n/2:n/2$ for $n$ even or $(n-1)/2:(n+1)/2$ for $n$ odd. It follows that a topology that is more balanced will typically receive more prior mass than a topology that is more unbalanced. In the remainder of this subsection we therefore use simulation to examine posterior robustness in cases where prior-likelihood conflict arises due to a data generating tree that is unbalanced or that contains a long branch.

\begin{figure*} 
\centering
\includegraphics[scale=0.5]{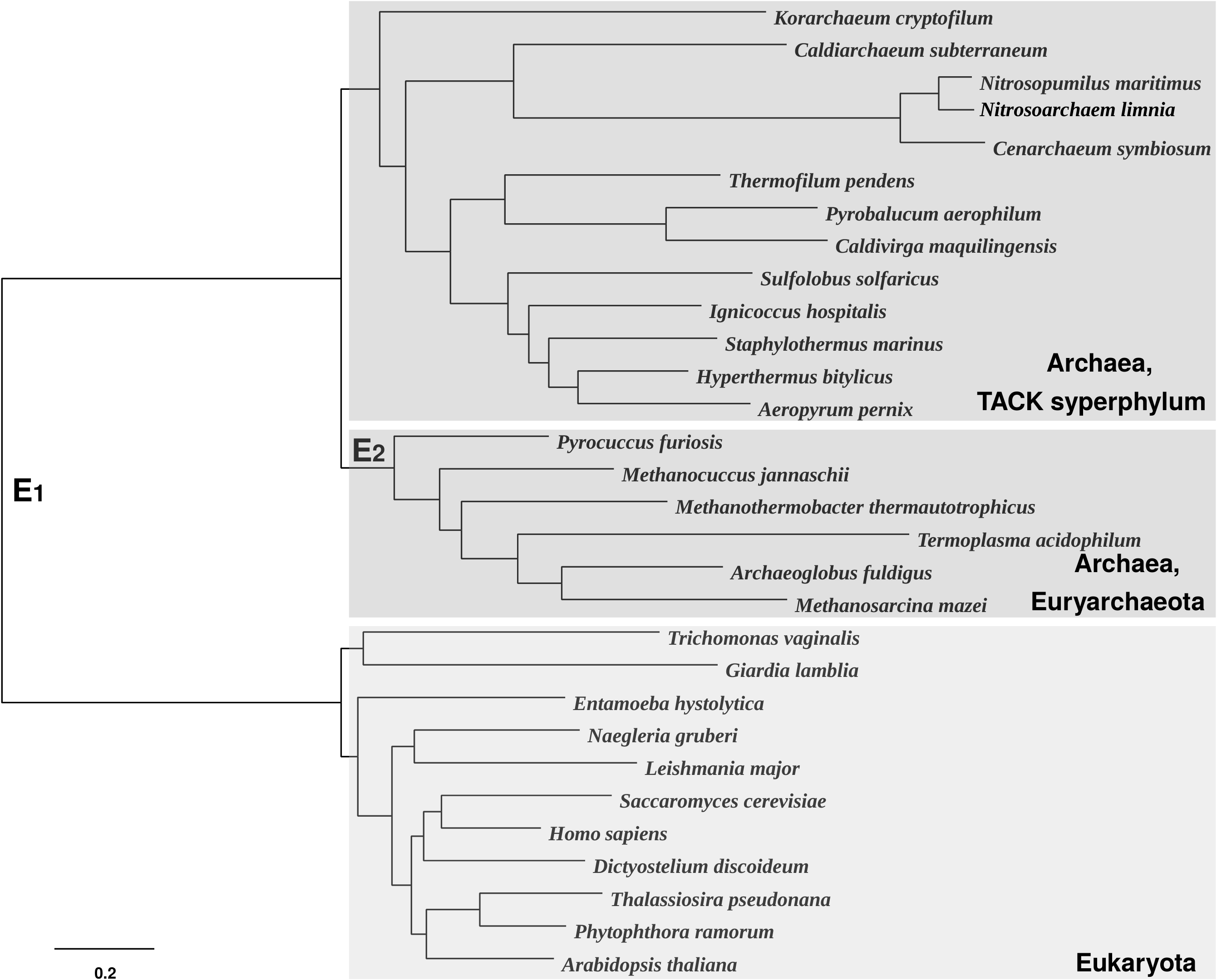}\newline
\caption{An unrooted 30-taxon tree derived from a recent analysis \citep{williams2012} describing the relationships between Archaea and Eukaryota. A root on the branch $E_1$ corresponds to the three-domains hypothesis (located between monophyletic Archaea and Eukaryota), while a root on the branch $E_2$ corresponds to the eocyte hypothesis (located within  paraphyletic Archaea, separating Euryarchaeota from the clade comprising the TACK superphylum and Eukaryota).}
\label{fig:AEunrootedTree}
\end{figure*}

We base our simulations on an unrooted 30-taxon tree derived from a recent analysis (Figure \ref{fig:AEunrootedTree}) \citep{williams2012}.
This tree describes the relationships between Archaea and Eukaryota. 
These relationships are still debated, concentrating on two competing hypotheses about the tree of life: (i) the three-domains hypothesis, according to which the root of the tree comprising Archaea and Eukaryota is placed on the branch separating monophyletic Archaea from monophyletic Eukaryota (branch $E_1$), and (ii) the eocyte hypothesis which places the root within a paraphyletic Archaea (branch $E_2$). Based on this unrooted tree, we construct six different rooted trees by changing the placement of the root and the length of the branch $E_1$ according to Table \ref{tab:trees}.

\begin{table}
\begin{center}
\caption{Six rooted trees for simulating the data. The trees have the unrooted topology of the tree depicted in Figure \ref{fig:AEunrootedTree} but differ in the placement of the root and the length of the branch $E_{1}$. Note that if a tree is rooted on branch $E_i$, the root is placed at the middle of $E_i$. }
\label{tab:trees}
\begin{tabular}{ c    c  c }
   Tree &Root edge & Length of $E_{1}$ \\ \hline                                      
   1 &$E_{1}$ & 1.3 \\ 
   2 &$E_{2}$ & 1.3 \\
   3 &$E_{1}$ & 0.1 \\
   4 &$E_{2}$ & 0.1 \\
   5 &$E_{1}$ & 0.3 \\
   6 &$E_{2}$ & 0.3                 
\end{tabular}
\end{center} 
\end{table}

Trees 1, 3 and 5 are fairly balanced with root type $11:19$, whilst Trees 2, 4 and 6 are more unbalanced with root type $6:24$. The Yule prior assigns almost 30\% more mass to the former rooted topology. In Trees 1 and 2 and, to a lesser extent, Trees 5 and 6, the unrooted topology contains a long internal branch. In Trees 3 and 4 this internal branch is short. Given the unrooted tree depicted in Figure~\ref{fig:AEunrootedTree}, the prior will therefore support placement of the root on branch $E_1$, particularly if this branch is long.

We use the NR model to simulate a rate matrix $Q$ with $\boldsymbol\pi = (0.25, 0.25, 0.25, 0.25)$, $\kappa =2$ and $\sigma=0.3$. In turn, this rate matrix is used to simulate three different alignments for each tree. These alignments are then analysed under the NR model with the Yule prior.

\begin{figure*} 
\centering
\subfloat[][Tree 1.]{\label{fig:rootsTree1} \includegraphics[scale=0.7]{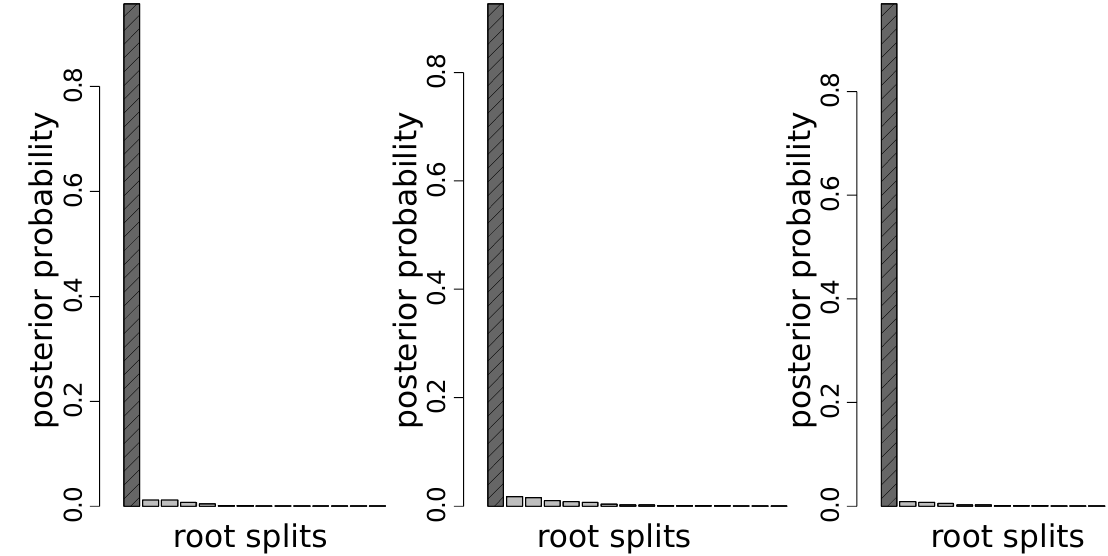}}  \hspace*{50.0pt} 
\subfloat[][Tree 2.]{\label{fig:rootsTree2} \includegraphics[scale=0.7]{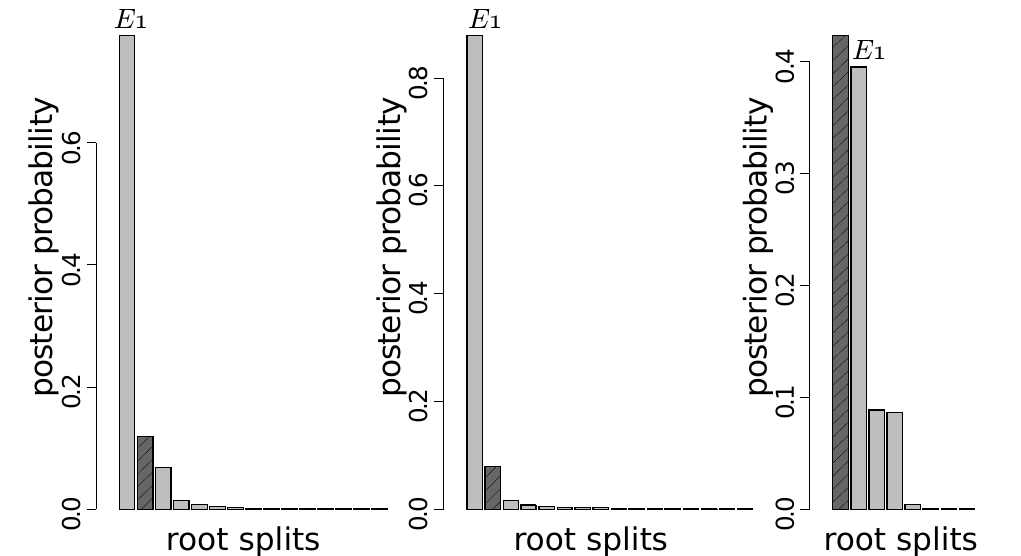}}\\
\subfloat[][Tree 3.]{\label{fig:rootsTree3} \includegraphics[scale=0.7]{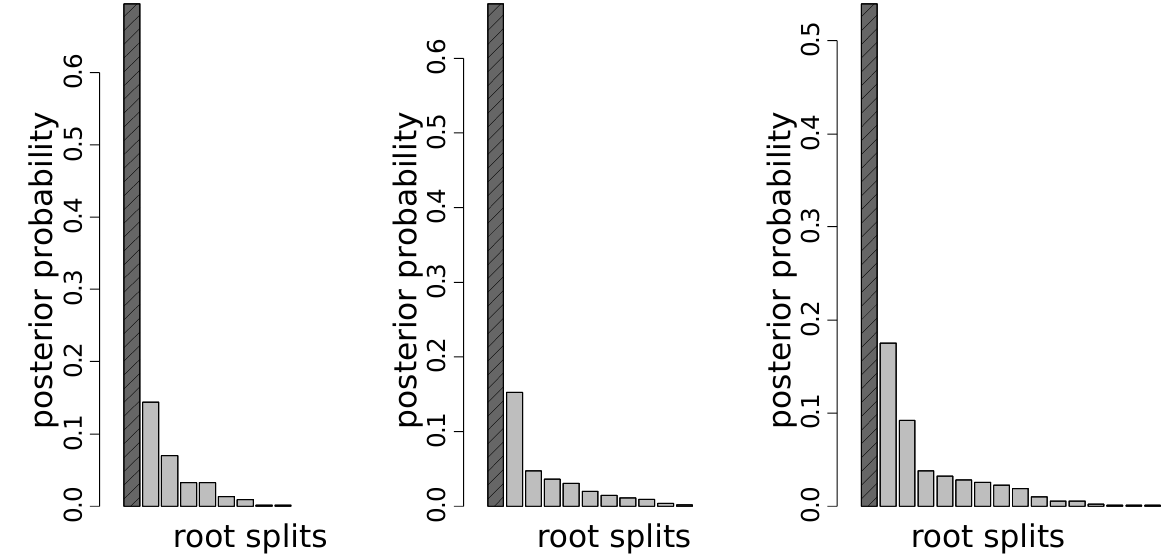}} \hspace*{40.0pt}
\subfloat[][Tree 4.]{\label{fig:rootsTree4} \includegraphics[scale=0.7]{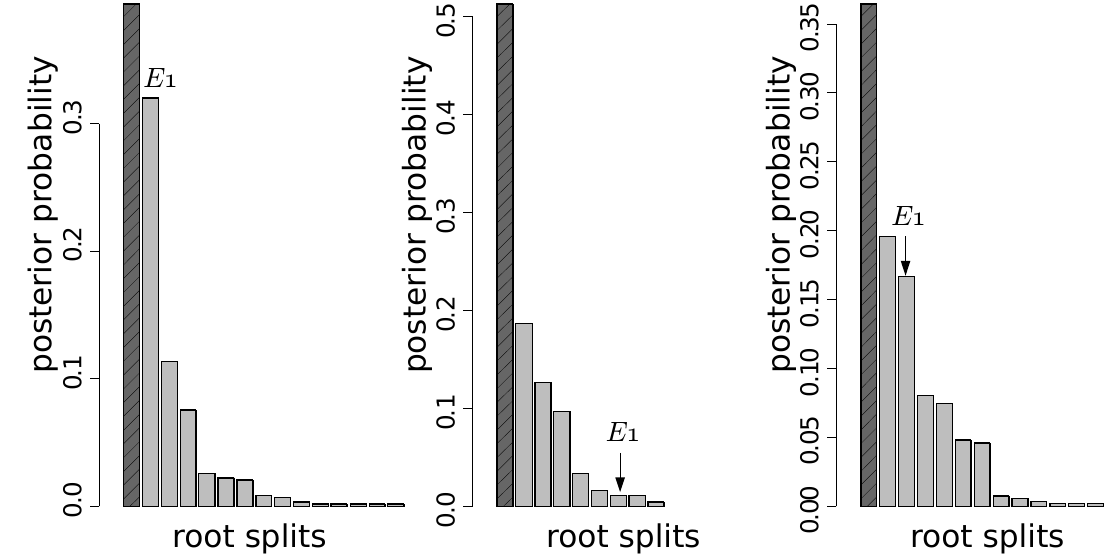}}\\
\subfloat[][Tree 5.]{\label{fig:rootsTree5} \includegraphics[scale=0.7]{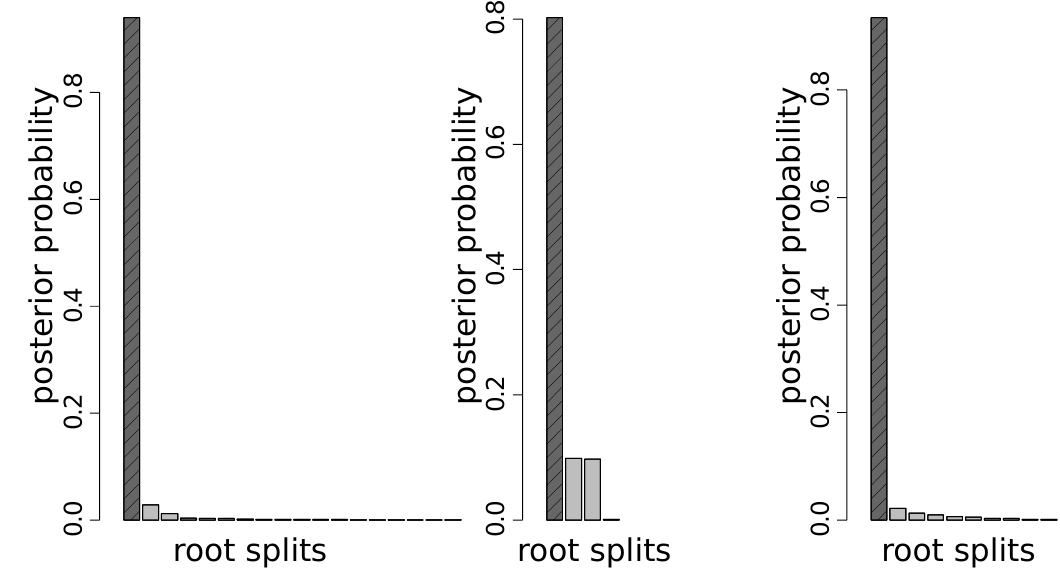}}\hspace*{60.0pt}
\subfloat[][Tree 6.]{\label{fig:rootsTree6} \includegraphics[scale=0.7]{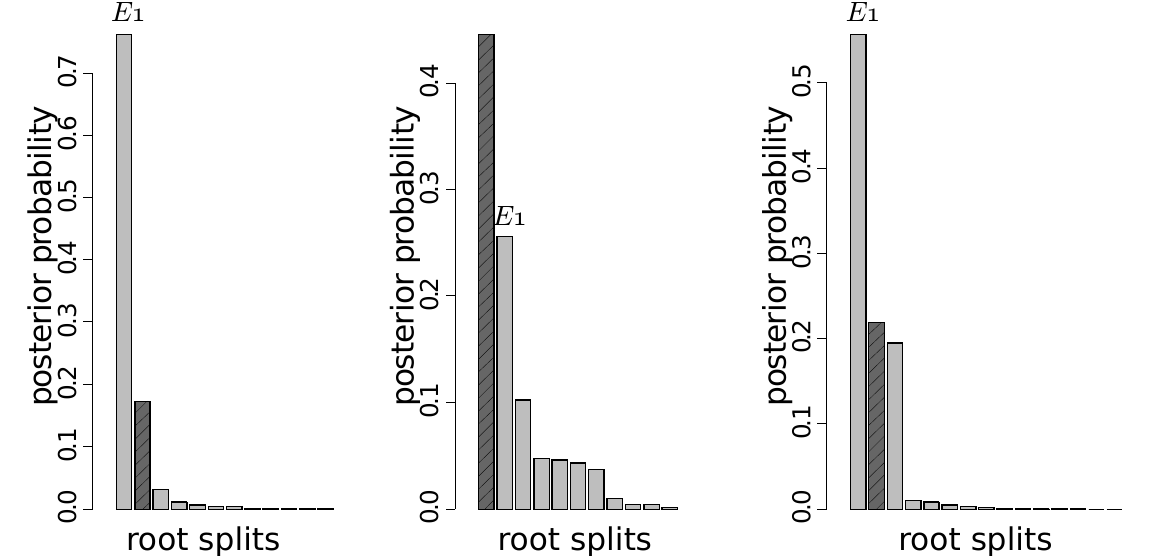}}
\caption{Posterior distribution of the root splits for three different alignments simulated for each of the six rooted trees according to Table \ref{tab:trees} . Different bars on each plot represent different root splits ordered by posterior probabilities, with the highlighted bar representing the true root split. In the plots for Trees 2, 4 and 6, the split corresponding to a root on edge $E_1$ is also marked.}
\label{fig:trees}
\end{figure*}

\begin{description}
\item[Tree 1:] Tree 1 is rooted on the long branch $E_1$. Clearly the likelihood for data generated from this tree will support the correct placement of the root. Moreover, for the reasons expressed above, the prior will also support rooting on edge $E_1$. It is not surprising, therefore, that we find the posterior is very concentrated around the true root position (Figure \ref{fig:rootsTree1}).
\item[Tree 2:] In Tree 2, the root is placed on the much shorter branch $E_2$, creating a fairly unbalanced unrooted topology with a long interior branch $E_1$. As such, data generated under this tree will favour the correct root position on edge $E_2$, but the prior will favour a root on branch $E_1$. This creates prior-likelihood conflict. As expected, we find that the posterior probability of the true root drops substantially in comparison to the analysis for Tree 1 and in two of the three analyses, the posterior offers more support to a root on edge $E_1$ (Figure \ref{fig:rootsTree2}).
\item[Tree 3:] Tree 3 has the same rooted topology as Tree 1 but the root branch $E_1$ is now much shorter and the unrooted topology does not contain any long edges. As for Tree 1, prior-likelihood conflict does not arise but there is no longer such pronounced prior support for placement of the root on edge $E_1$. Nevertheless, we find that the posterior is still concentrated around the true root position (Figure \ref{fig:rootsTree3}). 
\item[Tree 4:] Tree 4 has the same rooted topology as Tree 2 but the long interior branch $E_1$ is now shortened to 0.1. Although the Yule prior generally favours more balanced trees than Tree 4, the prior for branch lengths no longer offers overwhelming support to placement of the root on edge $E_1$. We find that the true root can now be recovered as the posterior mode (Figure \ref{fig:rootsTree4}) but with less support than in the analysis for Tree 3.
\item[Tree 5:] Tree 5 has the same rooted topology as Trees 1 and 3, but the root edge $E_1$ has length 0.3, which lies between the corresponding values for Trees 1 and 3. As expected, we find that the true root is inferred as the posterior mode (Figure \ref{fig:rootsTree5}), and the posterior is less (more) concentrated around the mode in comparison to the analysis of Tree 1 (Tree 3).
\item[Tree 6:] Tree 6 has the same rooted topology as Trees 2 and 4, but the internal edge $E_1$ has length 0.3, which lies between the corresponding values for Trees 2 and 4. The unrooted topology has a moderately long interior edge and the rooted topology is unbalanced, leading to some prior-likelihood conflict. We find that a root on edge $E_1$ sometimes receives more posterior support than the true root (Figure \ref{fig:rootsTree6}), although, as expected, this effect is less pronounced than in the analysis for Tree 2.
\end{description}

This simulation experiment illustrates the sensitivity of root inferences to conflict between the prior and the likelihood. The effect of a mismatch in information about branch lengths is particularly noticeable. Given a particular unrooted topology, whilst the likelihood might support the presence of a long branch in the corresponding rooted tree, an Exp(10) prior does not, and therefore favours placement of the root on the long edge. Ideally constructing a more flexible prior that more explicitly models topology and branch lengths jointly will contribute to better root inference. However, given the absence of very long branches, our results show that the model is still able to extract information from the data about the root even in the face of prior-likelihood conflict.

\subsection{Analysis of experimental data}

\subsubsection{Rooting the radiation of palaeopolyploid yeasts}

\begin{figure*}
\centering
	\includegraphics[scale=0.6]{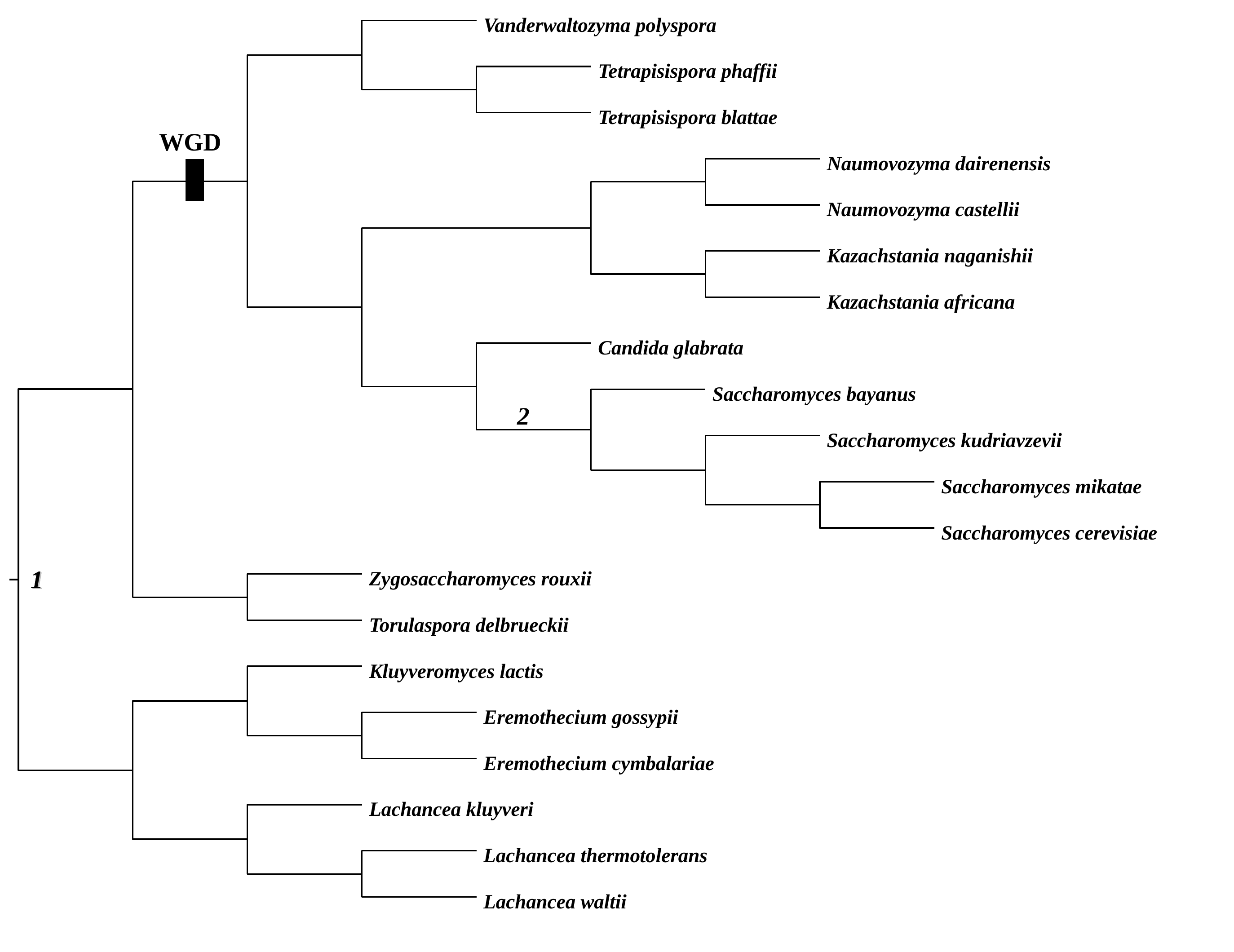}
\caption{Rooted phylogeny of the palaeopolyploid yeasts supported by the whole-gene duplication analysis (not drawn to scale), reproduced from the YGOB website \citep{byrne2005, ygob}. The tree is rooted according to the outgroup method based on an analysis with the GTR+I+G model in a maximum likelihood framework \citep{hedtke2006}. Roots 1 and 2 represent the two most plausible posterior root splits in the current analysis. }
\label{fig:yeastTrueTree}
\end{figure*}

We next investigated the performance of the NR and NR2 models on a real biological data set for which there is broad biological consensus on the root position \citep{byrne2005, hedtke2006}. The lineage leading to \emph{Saccharomyces cerevisiae} (brewer's yeast) and its relatives underwent a conserved whole-genome duplication (WGD) about 100 million years ago \citep{ wolfe_shields1997, lander2004}. Evidence for this WGD, in the form of duplicated genes and genomic regions, is shared by all post-WGD yeasts and defines the group as a clade from which the root of the \emph{Saccharomycetales} is excluded (Figure \ref{fig:yeastTrueTree}) \citep{byrne2005, ygob}. 

\begin{figure*}
\centering
\subfloat[][]
	{\label{fig:rootsYeastSu}\includegraphics[scale=0.45]{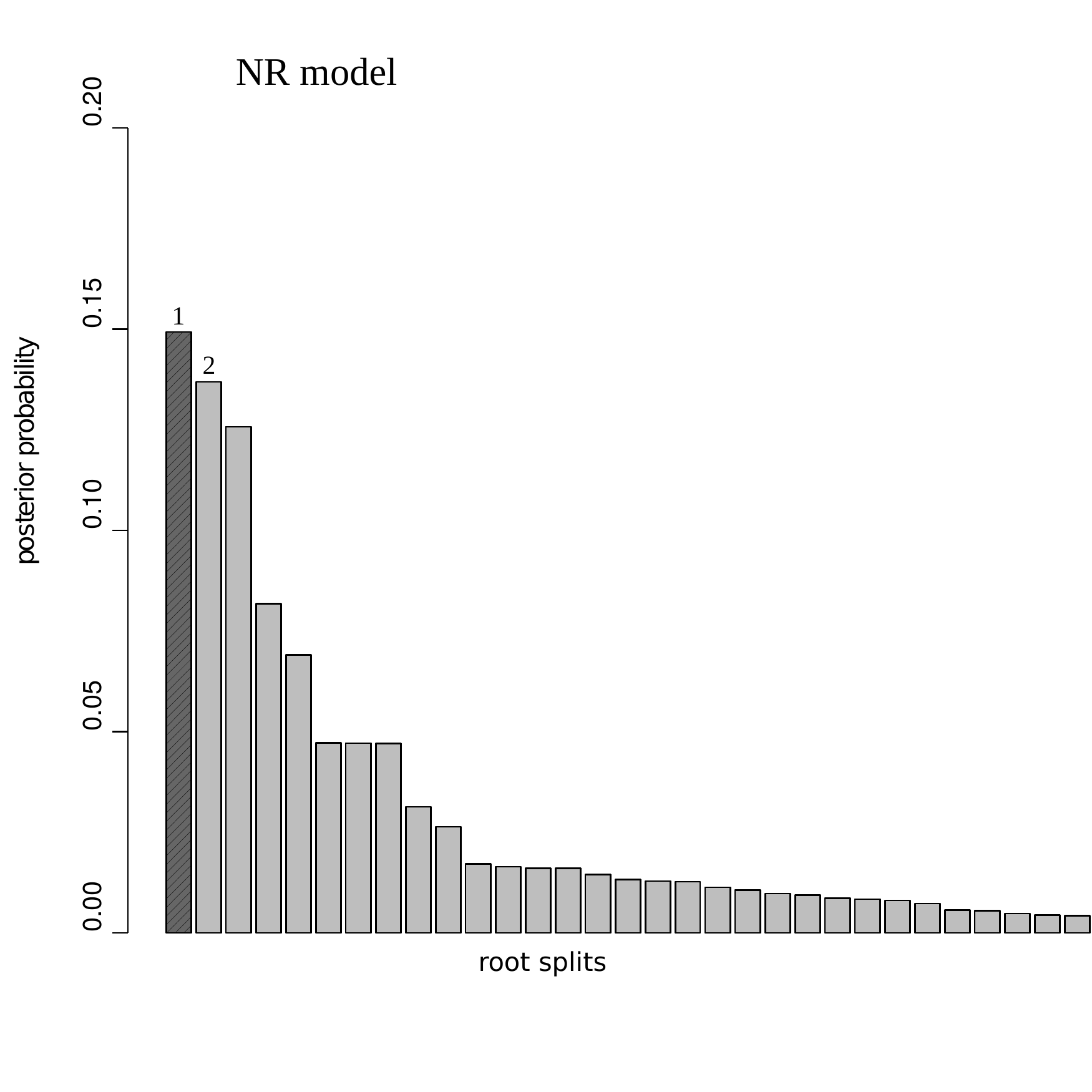}
	\includegraphics[scale=0.45]{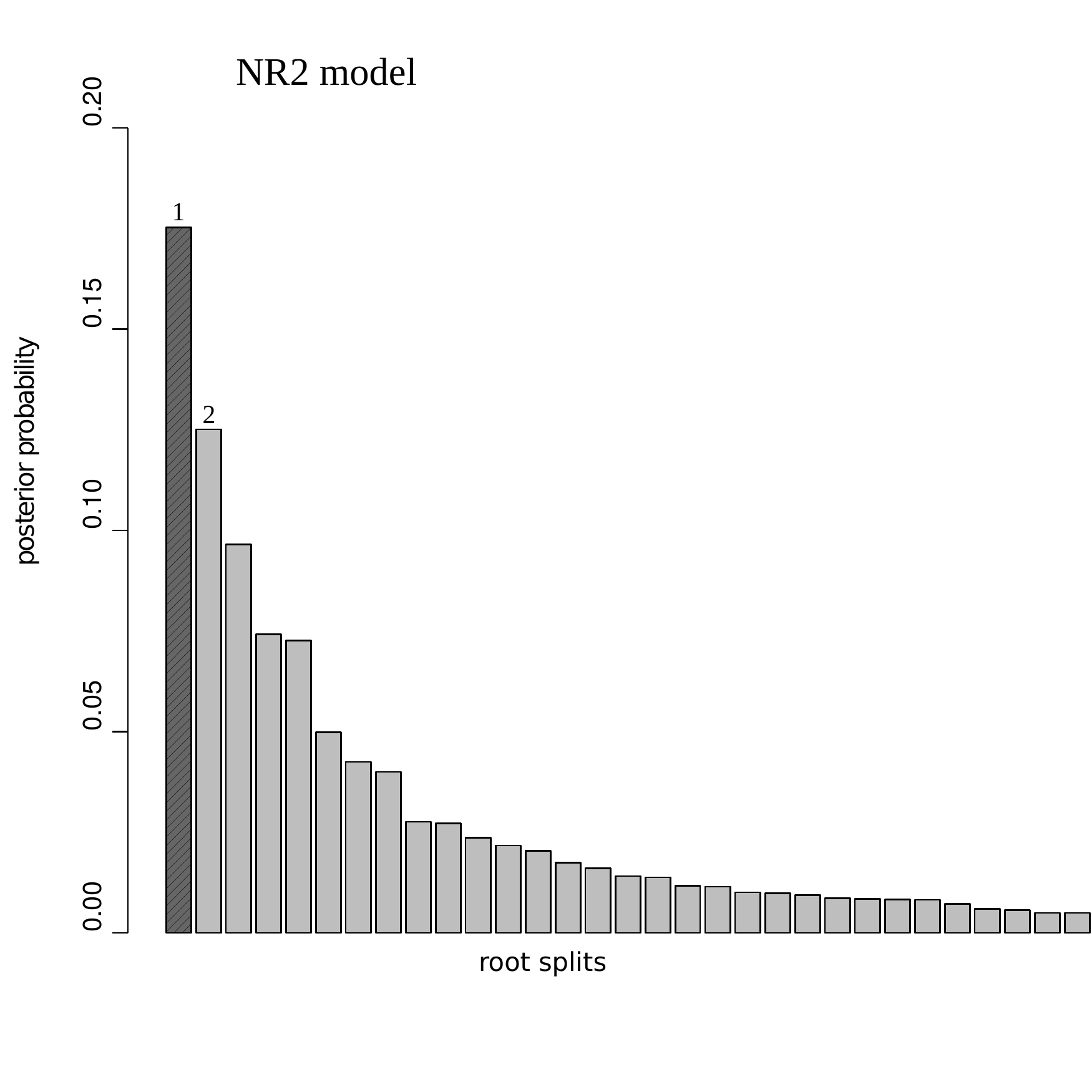}}\\
\subfloat[][]	
	{\label{fig:rootsYeastYule}\includegraphics[scale=0.45]{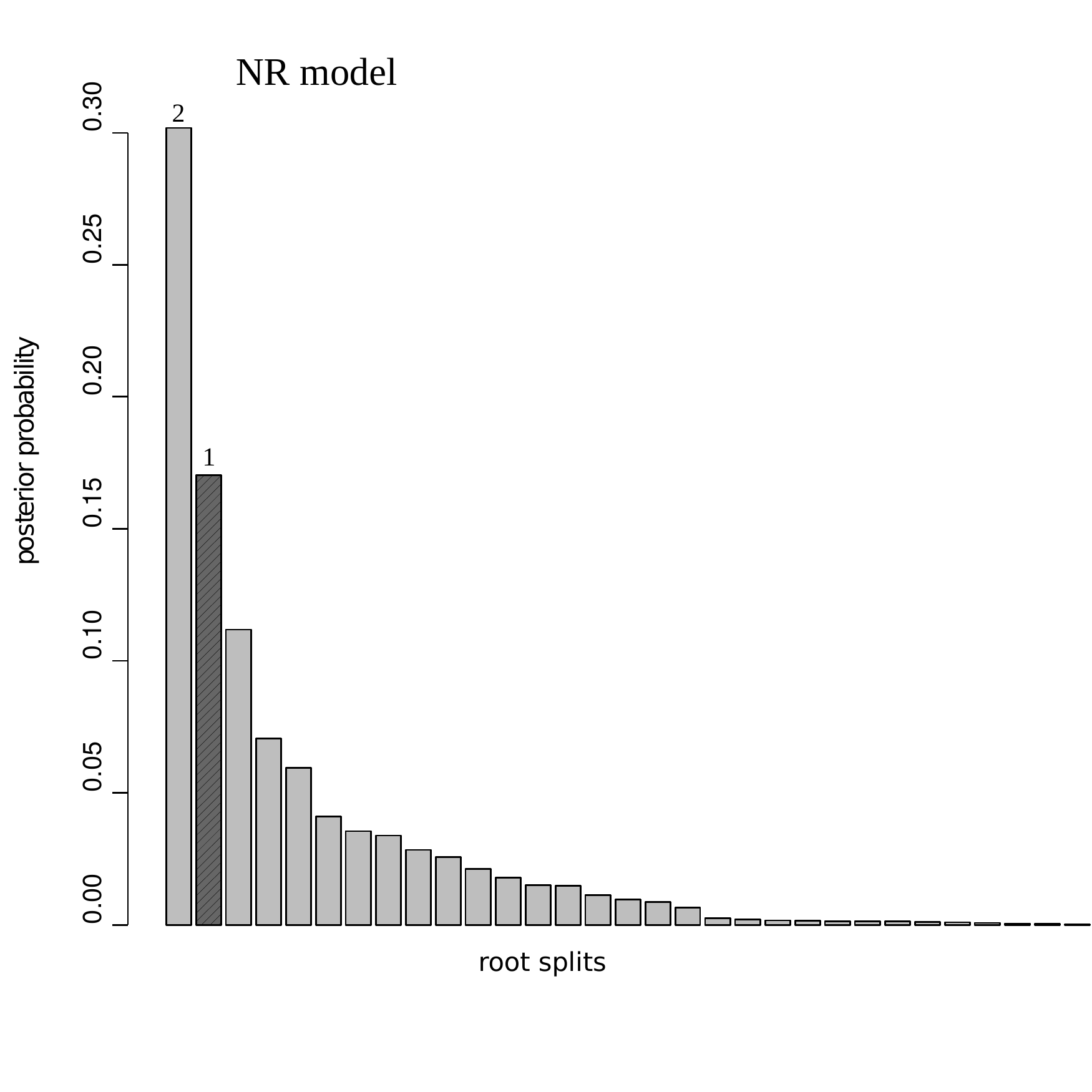}
	\includegraphics[scale=0.45]{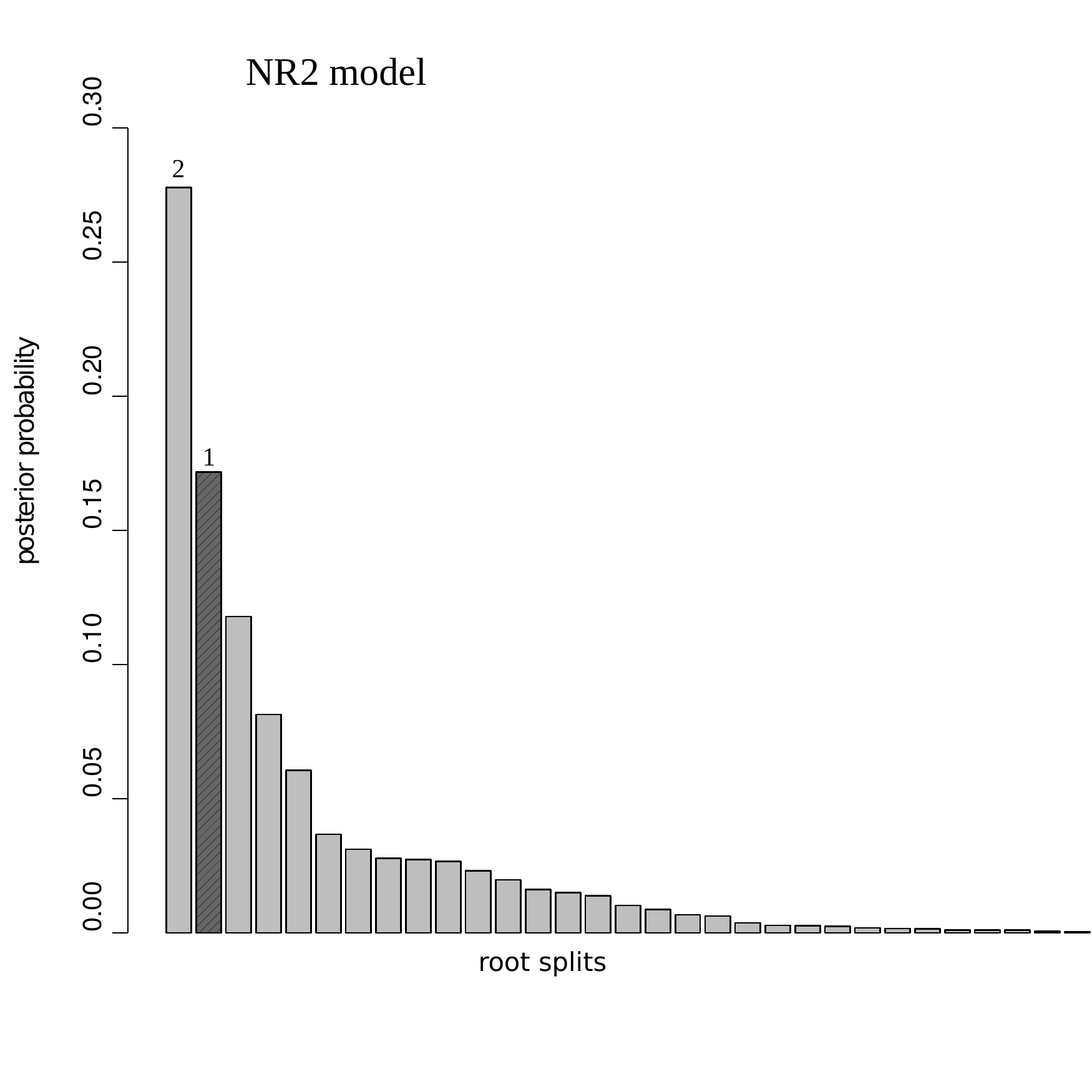}}
\caption{The posterior distribution of the root splits of the palaeopolyploid yeasts data set for both NR and NR2 models analysed (a) with the structured uniform prior and (b) with the Yule prior. Different bars on the plot represent different root splits on the posterior distribution of trees ordered by posterior probabilities (roots 1 and 2 are mapped in Figure \ref{fig:yeastTrueTree}). In (a), the analysis performed with the structured uniform prior, the root split supported by outgroup rooting (Hedtke at al. 2006) has the highest posterior probability (root 1, highlighted), while root 2 is placed within the post-WGD clade. In (b), the analysis performed with the Yule prior, the root split supported by outgroup rooting (Hedtke at al. 2006) has the second highest posterior probability (root 1, highlighted). The posterior modal root 2 is placed within the post-WGD clade.}
\label{fig:rootsYeast}
\end{figure*}

The root inferred through outgroup analysis separates a clade comprising \textit{Eremothecium gossypii, Eremothecium cymbalariae, Kluyveromyces lactis, Lachancea kluyveri, Lachancea thermotolerans} and  \textit{Lachancea waltii} from the other species \citep{hedtke2006}.       
We analysed an alignment of concatenated large and small subunit ribosomal DNA sequences for 20 yeast species, with a combined length of 4460 bp. The sequences were aligned with MUSCLE \citep{edgar04}, and poorly aligned regions were detected and removed using TrimAl \citep{capella09}.  
We analysed this data set with the NR and NR2 models, using both the Yule prior and the structured uniform prior. In the analysis with the structured uniform prior, the root split supported by outgroup rooting \citep{hedtke2006} has the highest posterior probability (root 1 in Fig. \ref{fig:yeastTrueTree}) for both models. However, there is a substantial amount of uncertainty represented by the non-negligible posterior probabilities of the other root splits (Fig. \ref{fig:rootsYeastSu}) and, for example, 
the second most plausible root is located within the post-WGD clade (root 2 in Fig. \ref{fig:yeastTrueTree}).
This posterior uncertainty is also reflected in the sensitivity of the analysis to the topological prior: while the structured uniform prior recovered the root supported by the outgroup analysis with the highest posterior support, the Yule prior instead recovered this root with the second-highest support (Fig. \ref{fig:rootsYeastYule}). The most plausible root inferred with the Yule prior is placed within the post-WGD clade (root 2 in Fig. \ref{fig:yeastTrueTree}) contradicting the WGD analysis. 

The posterior for Huelsenbeck's $I$ statistic is suggestive of a non-negligible degree of non-reversibility in the data (the posterior mean is 0.2 for the analysis with the NR model, 0.14 for the analysis with the NR2 model). In our simulations, larger values of $I$ were generally required to infer the true root with high posterior probability. However, the support offered to the widely accepted outgroup root in this analysis shows that it is possible to extract useful root information in spite of the data suggesting only a modest degree of non-reversibility.

\begin{figure*}
\centering
\subfloat[][]	
	{\label{fig:consYeastSu}\includegraphics[scale=0.48]{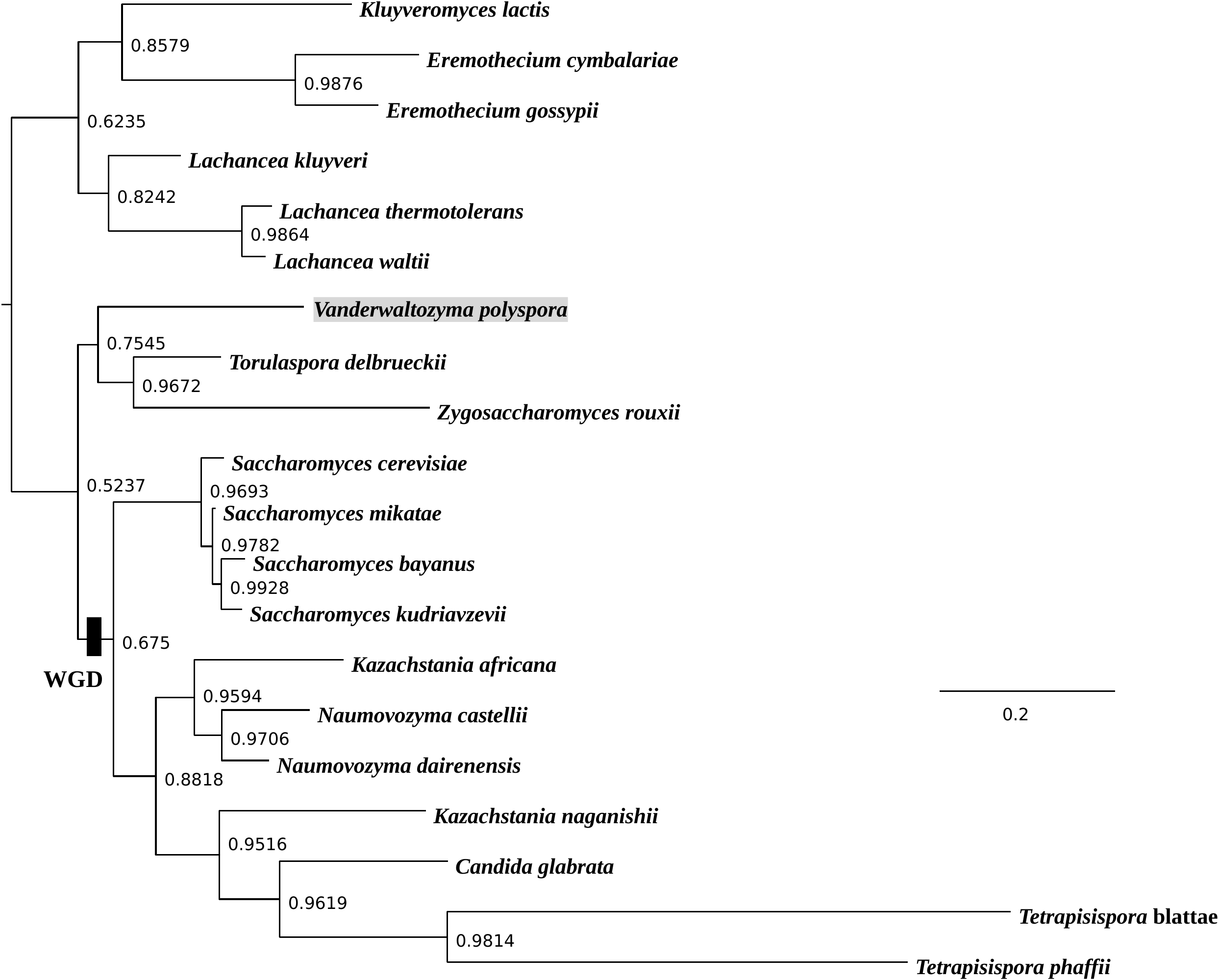}}\\
\subfloat[][]	
	{\label{fig:consYeastYule}\includegraphics[scale=0.48]{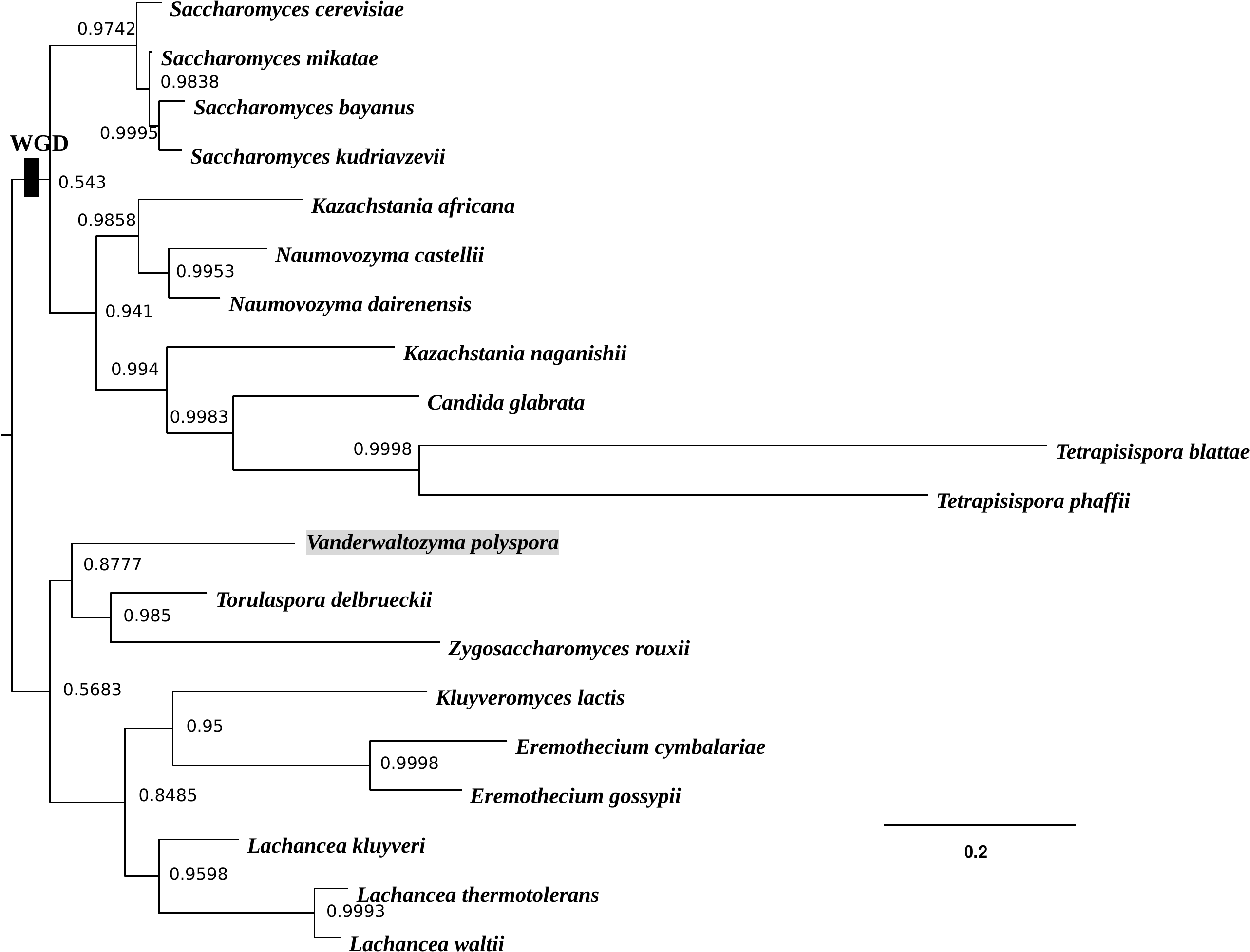}}
\caption{Rooted majority rule consensus tree of the palaeopolyploid yeasts data set, inferred under the NR model using (a) the structured uniform prior and (b) the Yule prior, with the WGD event mapped. The analysis is based  on the alignment of concatenated large and small subunit ribosomal DNA sequences for 20 yeast species, 4460 bp. The trees differ from that supported by the WGD analysis by the placement of \emph{Vanderwaltozyma polyspora} (highlighted) within the pre-WGD clade.}
\label{fig:consYeast}
\end{figure*}

The unrooted topologies of the rooted majority rule consensus trees from the analyses with the two topological priors (Fig. \ref{fig:consYeast}) differ from that supported by the WGD analysis by the placement of \emph{Vanderwaltozyma polyspora}. While the WGD analysis places it within the post-WGD clade, in our analysis this taxon is located within the pre-WGD clade.
This result is consistent with our posterior inferences from fitting the HKY85 and GTR models. Interestingly, it is also consistent with the analysis performed with the site-heterogeneous CAT-GTR model \citep {lartillot2004} where  \emph{Vanderwaltozyma polyspora} is, again, excluded from the post-WGD clade (not shown).
The placement of \emph{Vanderwaltozyma polyspora} outside the WGD clade is surprising given that the genome of \emph{Vanderwaltozyma polyspora} preserves evidence of having undergone WGD \citep{scannell2007}. While this result requires further investigation, the similarity between the consensus trees obtained with the CAT-GTR model and with our non-reversible models suggests that the non-reversible models can not only extract meaningful information about the root position, but also capture information for inferring the unrooted topology. However, the minor mismatch of the topologies inferred in our analysis with that supported by WGD and outgroup analyses \citep {hedtke2006} confirms the presence of some features of the data that our models do not account for. For example, ribosomal RNA function depends on the molecule folding into a complex three-dimensional shape. Interactions among sites that are distant in the primary sequence, but close in the three dimensional structure, are likely to induce site-specific selective constraints that are not accounted for in our models. In addition, it has been previously shown that failure to account for compositional heterogeneity can lead to inferring incorrect topologies with strong support \citep{foster2004, cox, foster2009, williams2012}.
Thus further refinement of the models, for instance, relaxing the stationarity assumption, might be necessary to improve the ability of the models to provide better insight into the evolution of palaeopolyploid yeasts.\par
It is worth noting that the root split on the majority rule consensus tree (Fig. \ref{fig:consYeastYule}) does not match the marginal posterior modal root split (Fig. \ref{fig:rootsYeastYule}). This happens because the consensus tree is a conditional summary, computed recursively from the leaves to the root, which depends upon the plausibility of subclades. On the other hand, the posterior over root splits is a marginal summary that averages over the relationships expressed elsewhere in the tree; see Appendix 2 for an illustrative example. 

\subsubsection{Analysis of the ribosomal tree of life}

We have also applied the models to a data set for which there is still debate about the unrooted topology and root position: the ribosomal tree of life. Recall that the debates are centred on two hypotheses. According to the three-domains hypothesis,  Archaea is monophyletic, sharing a common ancestor with Eukaryota \citep{woese}.
The other hypothesis, called the eocyte hypothesis, suggests that Archaea is paraphyletic and Eukaryota originated from within Archaea \citep{lake_88, rivera_lake_92, cox}. 
Recent analyses of ribosomal RNA data have demonstrated that topological inferences can be sensitive to the choice of substitution model. When homogeneous models are used for the analysis they often recover the three-domains tree, while heterogeneous models generally recover the eocyte tree \citep{cox, williams2012}. In addition, there is also external evidence for the eocyte hypothesis. For example, newly discovered archaeal species whose genomes encode many eukaryote-specific features, provide additional support for the eocyte hypothesis \citep{spang2015}.
\begin{figure*}	[ht!]
	\hspace*{-45.0pt}	
	\includegraphics[scale=0.7]{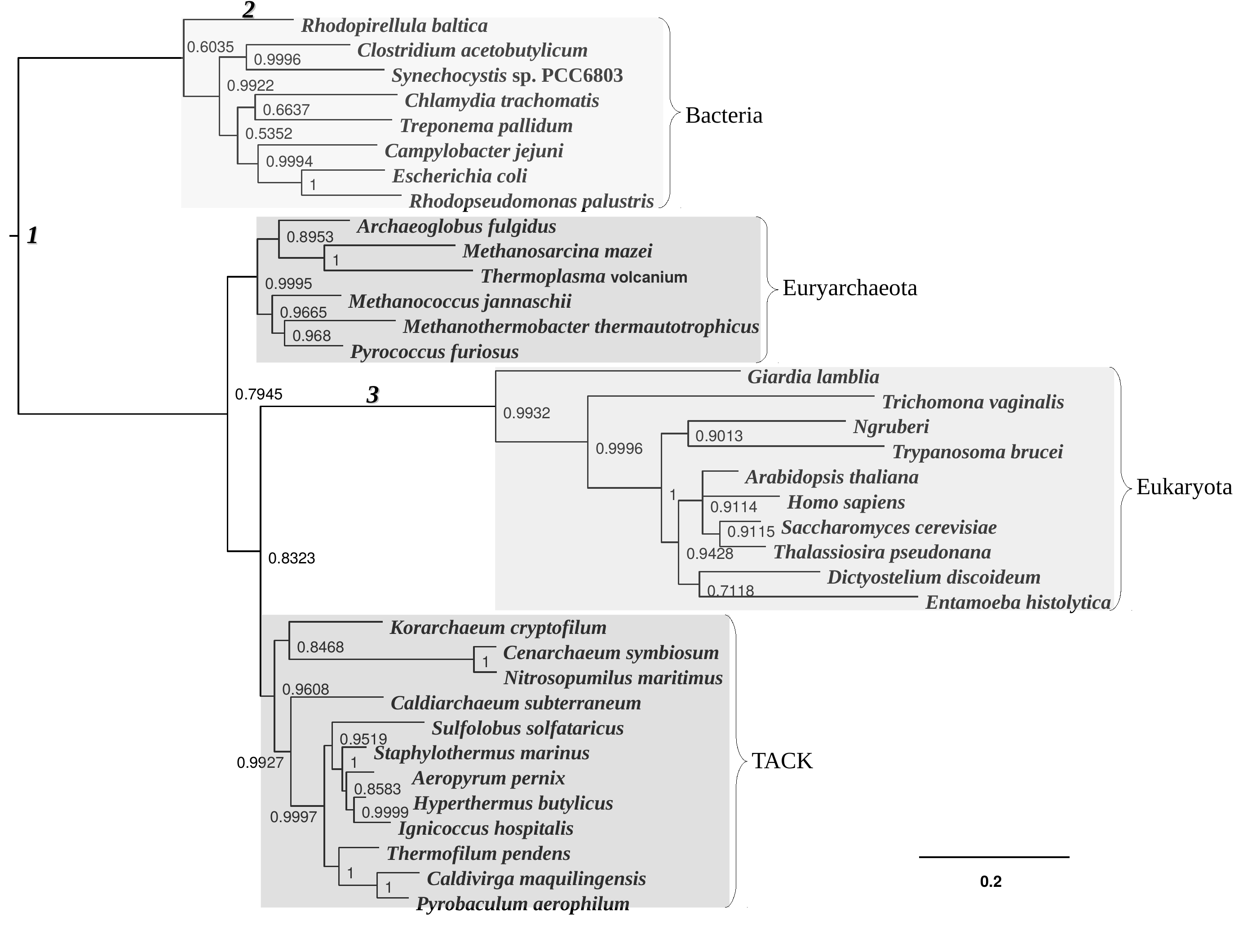}
\caption {Rooted majority rule consensus tree for the tree of life data set, inferred under the NR model using the Yule prior. The tree supports the eocyte hypothesis by placing Eukaryota within Archaea, as a sister group to the TACK superphylum. 
Roots 1, 2 and 3 are the root splits having the highest posterior support in the current analysis. Posterior support for these root splits is shown in Figure 7.}
\label{fig:tolCons}
\end{figure*}

\begin{figure*}
	\hspace{-30.0pt}
	\subfloat[][]
	{\label{fig:tolRootsNRYule} \includegraphics[scale=0.55]{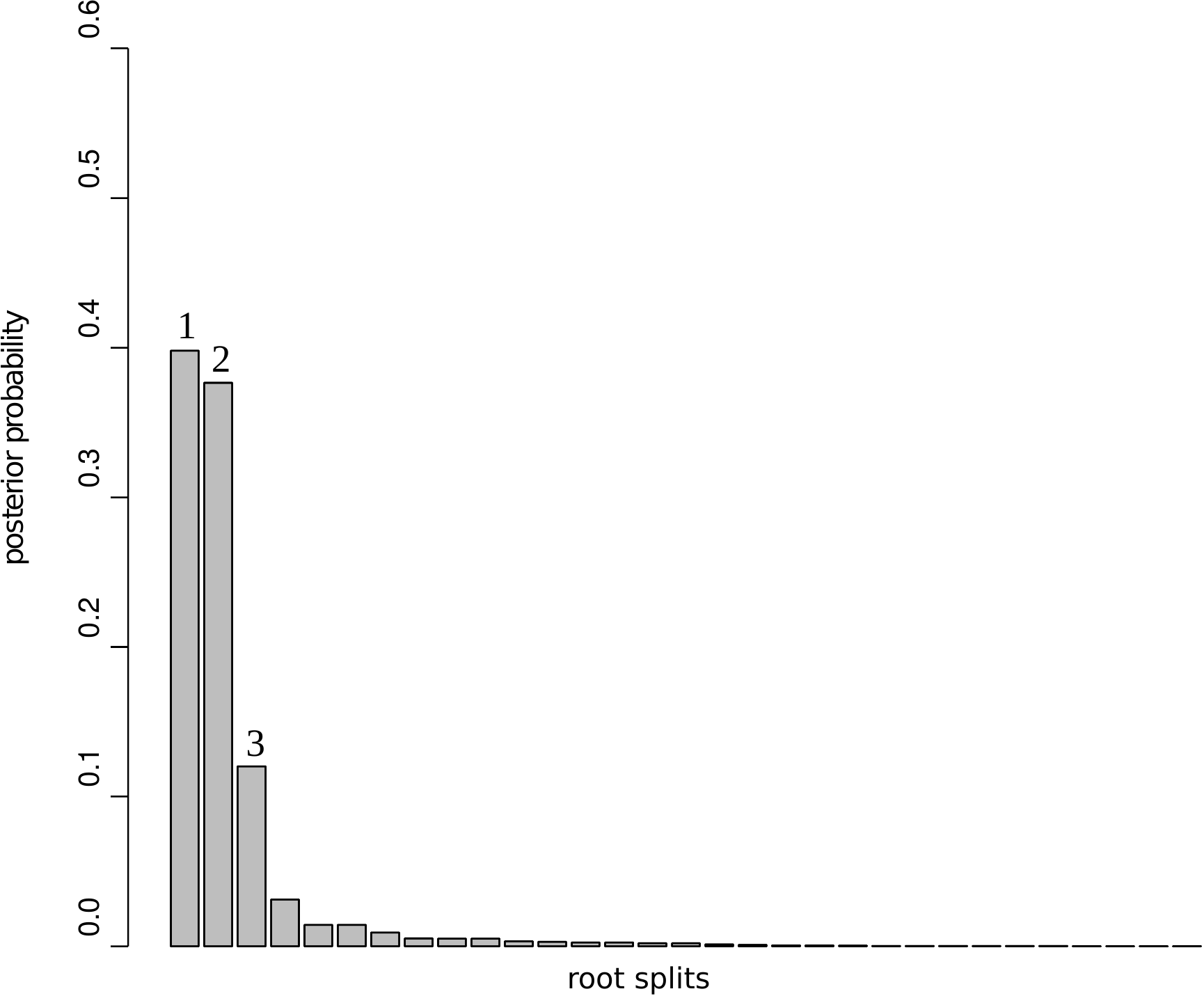}}
	\hspace{-30.0pt}	
    \subfloat[][]
	{\label{fig:tolRootsNRsu} \includegraphics[scale=0.55]{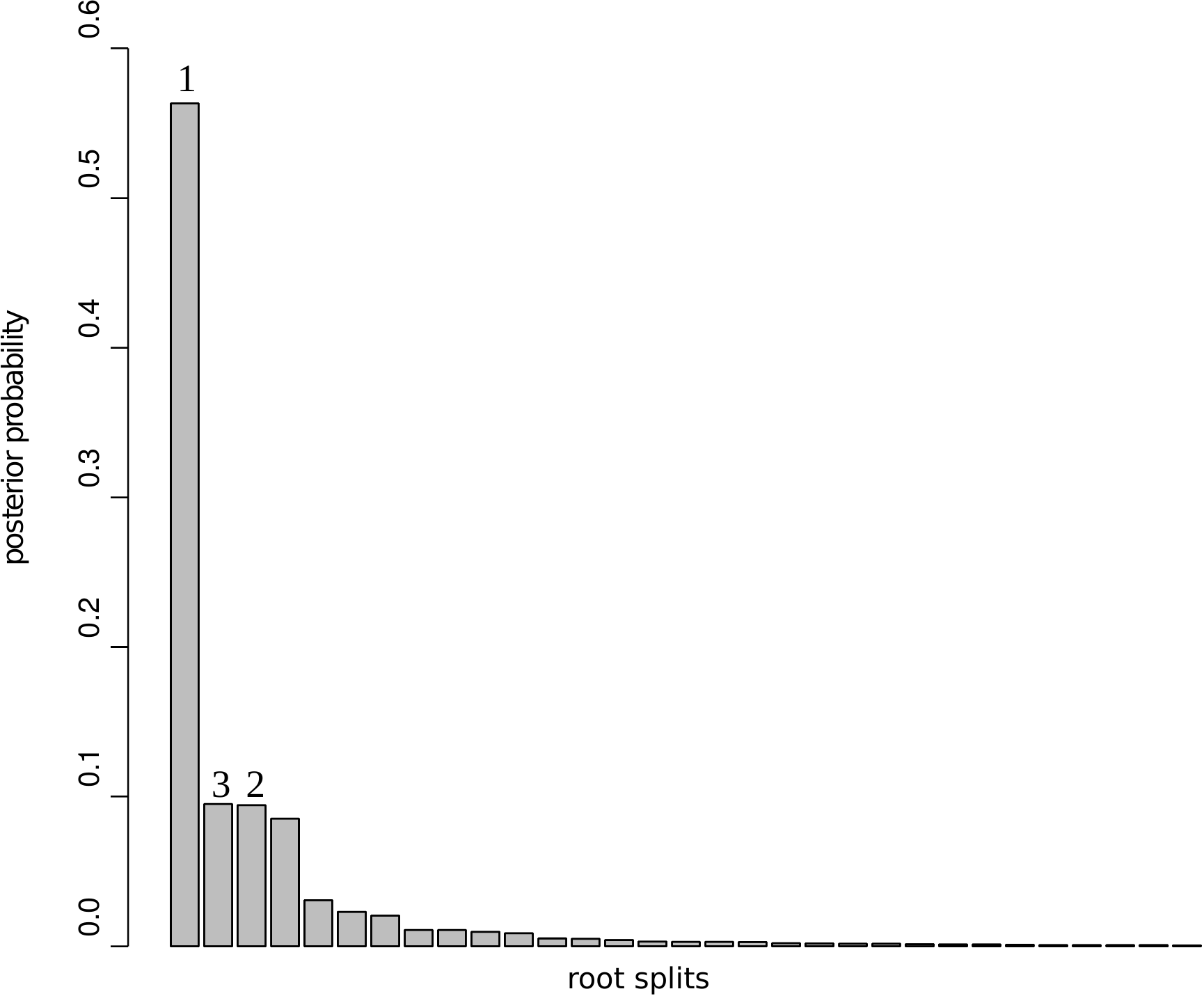}}
\caption{The posterior distribution of the root splits of the tree of life data set for the NR model analysed with (a) the Yule prior; and (b) with the structured uniform prior. Different bars on the plot represent different root splits on the posterior distribution of trees (ordered by posterior probabilities). The root split on the branch leading to Bacteria  has the highest posterior probability (root 1). Root 2 is placed within Bacteria (on the branch leading to \emph{Rhodopirellula baltica}) and root 3 is placed on the branch leading to  Eukaryota (the roots are mapped in Figure \ref{fig:tolCons}).}
\label{fig:tolRoots}
\end{figure*}

Here we analysed aligned concatenated large and small subunit ribosomal RNA sequences from archaeal, bacterial and eukaryotic species (36 taxa, 1734 sequence positions), including the recently discovered archaeal groups: Thaumarchaeota, Aigarchaeota and Korarchaeota. These new groups are closely related to Crenarchaeota and together they form the so-called TACK superphylum \citep {guy2011,kelly2011, williams2012, lasek2013}. Previous analysis of this data set performed with the CAT-GTR model recovered an eocyte topology \citep{williams2012}. Fitting the simpler HKY85 and GTR models also support this hypothesis. However these analyses were not able to infer the root because they used only reversible rate matrices in stationary substitution models. 
We analysed these data with both the NR and NR2 models using both the Yule prior and the structured uniform prior. In all cases we recovered the eocyte topology with similar posterior support (Fig. \ref{fig:tolCons}). The analysis with the Yule prior assigned high posterior support to two roots splits (Fig. \ref{fig:tolRootsNRYule}) - one on the branch leading to Bacteria (root 1 in Fig. \ref{fig:tolCons}), the other within Bacteria, on the branch leading to \emph{Rhodopirellula baltica} (root 2 in Fig. \ref{fig:tolCons}). This inference is in accord with current biological opinion about the root of the tree of life, which places the root either on the branch leading to Bacteria, or within  Bacteria \citep{baldauf, cavalier_smith, skophammer, hashimoto_hasegawa}. However, in  the analysis performed with the structured uniform prior, the support for the root within  Bacteria decreased and that for the the root on the bacterial branch increased (Fig. \ref{fig:tolRootsNRsu}). This analysis illustrates the sensitivity of the inference to the choice of topological prior, and confirms the importance of the choice of prior in Bayesian phylogenetics. The posterior mean of the Huelsenbeck's $I$ statistic is 0.18 for the analysis with the NR model and 0.17 for the analysis with the NR2 model. Again, this is suggestive of a moderate degree of non-reversibility in the data. Therefore, modelling other features of the data that also provide root information could make a  valuable contribution to the inference.

\section{Discussion}
We presented two hierarchical non-reversible models for inferring rooted phylogenetic trees.
The non-reversibility of both models is achieved by applying a stochastic perturbation to the rate matrix of a reversible model. This perturbation makes the likelihood dependent on the position of the root, enabling us to infer the root directly from the sequence alignment. 
In the first model (the NR model) we use only one variation component and perform a log-normal perturbation on the space of all possible rate matrices. In contrast, the second model (the NR2 model) utilises two variation components and the perturbation is performed on the space of reversible and non-reversible rate matrices separately. This separation allows us to judge the extent of the different types of perturbation.

The results on the simulated data with different levels of non-reversibility show that the correct root can be recovered with greater posterior support when the degree of non-reversibility in the data generating model is larger. We also investigated the robustness of posterior root inferences to situations where information from the prior and data are in conflict. Given a particular unrooted topology, our Yule prior for rooted trees and Exp(10) prior for branch lengths offers most support to balanced trees with short edges. Our simulations show that we can still recover the true root in the posterior when the data generating tree is unbalanced or the associated unrooted topology contains a long edge. However, when this edge is very long, it can mislead the root inference.  

We applied our models to two biological data sets. These analyses agree with our simulations in suggesting that our non-reversible models can recover useful rooting information, this time from real biological sequence alignments.
The analyses of both the yeast and tree of life data sets recover the widely agreed root. However, both data sets show some prior sensitivity, even though the two topological priors (the Yule prior and the structured uniform prior) share similar features. 
\label{pg:3.2}In order to investigate this issue we computed a log Bayes factor \citep{kass95} to compare the Yule prior ($Y$) with the structured uniform prior ($S$) for both examples with real data. Although usually used to compare models, the Bayes factor really compares prior-likelihood combinations and so can also be used to assess which of the two priors is most consistent with the data. The log Bayes factor for the yeasts data set is $\mathrm{log} B_{YS}$ = 2.27 suggesting that there is evidence against the structured uniform prior, however, the evidence is not strong. The log Bayes factor for the tree of life data set is $\mathrm{log} B_{YS}$ = 0.12 suggesting that there is no difference between the priors. Therefore the more noticeble prior sensitivity in the analysis of the yeasts data set is likely to be due to the greater difference in consistency between the data and each of the two priors.

Although Huelsenbeck's $I$ statistic provides evidence of a non-negligible  degree of non-reversibility in both biological data sets, the analyses display high levels of posterior uncertainty. This suggests that the information about the root may be obscured by other signals that are not accounted for by our current models.  For instance, our models assume the evolutionary process is stationary. If this was true then the empirical composition of the four nucleotides would be roughly the same for all taxa in the alignment. However, this is often not the case in experimental data \citep{foster2004,cox}. Notably, this assumption is violated for the tree of life data set where the empirical GC content ranged from 41\% for \textit{Entamoeba histolytica} to 69\% for \textit{Giardia lamblia}. The models may therefore benefit from further development, for example to model the non-stationarity of the process. Nonetheless, our findings illustrate that our non-reversible models NR and NR2 can be useful to infer the root position from real biological data sets. 

\section{Materials and methods}
We work within the Bayesian paradigm and base our inferences on the posterior distribution of the unknowns in the model. According to Bayes theorem, the posterior distribution is proportional to the prior times the likelihood. For the NR model, for example, the posterior distribution factorises as
\begin{equation}
\begin{split}
\pi(&\boldsymbol\pi, \kappa, \sigma, Q, \alpha, \boldsymbol\ell, \tau | \text{Data}) \propto \pi(Q|\boldsymbol\pi, \kappa, \sigma) \\ &\times \pi (\boldsymbol\pi, \kappa, \sigma, \alpha, \boldsymbol\ell, \tau) \times \pi(\text{Data}|Q, \alpha, \boldsymbol\ell, \tau).\label{eq:postNR}
\end{split}
\end{equation}
This distribution is analytically intractable and so we build up a numerical approximation by sampling from it using Markov chain Monte Carlo (MCMC) methods, specifically a Metropolis-within-Gibbs sampling scheme. In the remainder of this section, we first describe the calculation of the likelihood function, before outlining details of the MCMC algorithm. Finally, we provide practical details of the application of this algorithm to the analyses presented earlier in the \textit{Results} section.

\subsection{Likelihood}
The likelihood function summarises the information available from the data about the  unknowns in the model including the phylogenetic tree. Since we assume that alignment sites evolve  independently of each other, the likelihood can be expressed as a product of the likelihoods of the $n$ individual sites of the alignment. If we denote $\boldsymbol\theta$ to be the parameters of the substitution process, the likelihood takes the form  
\begin{equation*}
\pi(\text {Data}|\boldsymbol\theta, \alpha, \boldsymbol\ell, \tau) = \prod\limits_{i=1}^n\pi(D_i|\boldsymbol\theta, \alpha,  \boldsymbol\ell, \tau),
\end{equation*}
where $D_i$ is the column of nucleotides at site $i$. The probability of the data at a site $i$ is given by
\begin{equation*}
\pi(D_i| \boldsymbol\theta, \alpha, \boldsymbol\ell, \tau) = \sum\limits_{X}\pi_{X(root)} \prod\limits_{\text{edges} \hspace*{3.0pt} \ell = (v,w)} p_{X(v), X(w)}(\ell)
\end{equation*}
where $v$ and $w$ are the vertices at the two ends of edge $\ell$ and $X(u)$ denotes the nucleotide at a vertex $u$. The sum is taken over all functions $X$ from the vertices to $\Omega$ such that $X(u)$ matches data $D_i(u)$ for all leaf vertices $u$.\label{pg:1.6}
We assume a stationary model and so take the probability at the root $\pi_{X(root)}$ to be $\pi_{Q,X(root)}$, which comes from $\boldsymbol\pi_Q$, the theoretical 
stationary distribution associated with $Q$ (note that this is not the same as $\boldsymbol\pi$, the stationary distribution of the underlying HKY85 model).

\subsection{MCMC algorithm}
\subsubsection{NR model}
For the NR model, the posterior distribution for the unknowns in the model was summarised through equation~\eqref{eq:postNR}. At each iteration of the MCMC algorithm the following steps are performed: 
\begin{enumerate}[(a)]
\item update the parameters of the substitution model $(\boldsymbol\pi, \kappa, \sigma, Q, \alpha)$;
\item update the branch lengths $\boldsymbol\ell$ and the rooted topology $\tau$.
\end{enumerate}
In step (a) we update the parameters using a Dirichlet random walk proposal for $\boldsymbol\pi$ and log-normal random walk proposals for the other parameters. Move (b) consists of a series of Metropolis-Hastings steps to update each branch length one at a time using a log-normal random walk proposal and then updating the rooted topology and branch lengths (in a joint move) through three types of proposal: nearest-neighbour interchange (NNI), sub-tree prune and regraft (SPR), and a proposal that moves the root; see \citet{heaps2014} for complete details of all three moves. 

\subsubsection{NR2 model}
Here the posterior distribution of the unknowns takes the form
\begin{equation*}
\begin{split}
\pi(&\boldsymbol\pi, \kappa, \sigma_R, \sigma_N, \boldsymbol\epsilon, \boldsymbol\eta, \alpha, \boldsymbol \ell, \tau | \text{Data}) \\ &\propto \pi (\boldsymbol\pi, \kappa, \sigma_R, \sigma_N, \boldsymbol\epsilon, \boldsymbol\eta, \alpha, \boldsymbol \ell, \tau) \\ &\times  \pi(\text{Data}|\boldsymbol\pi, \kappa,  \boldsymbol\epsilon, \boldsymbol\eta, \alpha, \boldsymbol \ell, \tau)
\end{split}
\end{equation*}
and an analogous Metropolis-within-Gibbs algorithm is used to generate posterior samples.

\subsection{MCMC implementation}

In the \textit{Results} section, all results were based on (almost) un-autocorrelated posterior samples of size 5K. These samples were obtained by running the MCMC algorithm for at least 1000K iterations, discarding at least 500K iterations as burn-in and then thinning by taking every 100th iterate to remove autocorrelation. Convergence was diagnosed using the procedure described in \citet{heaps2014}. This involved initialising two MCMC chains at different starting points and graphically comparing the chains through properties based on model parameters and the relative frequencies of sampled clades. In all cases, the graphical diagnostics gave no evidence of any lack of convergence. The MCMC inferential procedures are programmed in Java and a software implementation can be found in the Supplementary Materials.

\section{Appendix 1}

The two-stage perturbation relies upon the underlying geometry of the space of Markov rate matrices, and is achieved in the following way. 
We work on a log-scale element-wise with all matrices, ignoring diagonal elements. 
The set of all possible $4 \times 4$ rate matrices $M$ is therefore identified with $\R^{12}$ which we equip with the standard inner product. 
The set of HKY85 matrices and GTR matrices form nested sub-sets of $M$. 
Recall that working element-wise on a log-scale, the off-diagonal elements of the rate matrix of the NR model can be expressed as, for $i \neq j$
\begin{equation} \label{eq:1}
\log q_{ij} =  \log q^{H}_{ij} + \epsilon_{ij},   
\end{equation}
where the $\epsilon_{ij}$ are independent $N(0,\sigma^2)$ quantities. The element-wise log of the HKY85 matrix $Q^H$ in equation~$\eqref{eq:1}$ is
\begin{equation}
\begin{split}
\log q^{H}_{ij} &= \tilde{\kappa}(\mathbf{e}_1\mathbf{e}_2^T+\mathbf{e}_2\mathbf{e}_1^T+\mathbf{e}_3\mathbf{e}_4^T+\mathbf{e}_4\mathbf{e}_3^T)+\sum_{i=1}^4 \tilde{\pi}_i\mathbf{s}\,\mathbf{e}_i^T\\
&= \tilde{\kappa}(\mathbf{e}_1\mathbf{e}_2^T+\mathbf{e}_2\mathbf{e}_1^T+\mathbf{e}_3\mathbf{e}_4^T+\mathbf{e}_4\mathbf{e}_3^T)
+\sum_{i=1}^3 \tilde{\pi}_i\mathbf{s}\,\mathbf{e}_i^T\\
&+\log\left( 1-e^{\tilde{\pi}_1}-e^{\tilde{\pi}_2}-e^{\tilde{\pi}_3} \right)\mathbf{s}\,\mathbf{e}^T_4 \label{eq:new}
\end{split}
\end{equation}
where $(\tilde{\pi}_1,\tilde{\pi}_2,\tilde{\pi}_3,\tilde{\pi}_4)$=$(\log\pi_A,\log\pi_G,\log\pi_C,\log\pi_T)$, $\tilde{\kappa}=\log\kappa$, $\std{i}$ is the $i$-th standard basis vector of $\R^4$ and $\mathbf{s}=(1,1,1,1)^T$. 
By differentiating~\eqref{eq:new} with respect to the parameters $\tilde{\pi}_1,\tilde{\pi}_2,\tilde{\pi}_3$ and $\tilde{\kappa}$ we obtain $4$ linearly independent vectors in $M$ that are locally tangent to the sub-set of HKY85 matrices at $Q^H$, and we denote these $V_1,V_2,V_3,V_4$. 
(Differentiating with respect to $\tilde{\pi}_4$ gives a tangent vector contained in the span of $V_1,V_2,V_3$.)
The tangent vectors in $M$ correspond to the $4\times 4$ matrices 
\begin{align*}
V_i &= \mathbf{s}\,\std{i}^T-\exp(\tilde{\pi}_i-\tilde{\pi}_4)\mathbf{s}\,\std{4}^T\quad \text{for\ }i=1,2,3,\\
\intertext{and\ }
V_4 &= \std{1}\std{2}^T+ \std{2}\std{1}^T+\std{3}\std{4}^T+ \std{4}\std{3}^T.
\end{align*}
The element-wise log of the general GTR matrix is 
\begin{equation*}
\sum_{i=1}^4\tilde{\pi}_i\mathbf{s}\,\std{i}^T+\sum_{i<j}\tilde{\rho_{ij}}\left( \std{i}\std{j}^T+ \std{j}\std{i}^T\right),
\end{equation*}
where $\tilde{\rho}_{ij}$ is the log of the exchangeability parameter $\rho_{ij}$.
\label{pg:1.15}By considering the derivatives with respect to the $\tilde{\rho}_{ij}$ parameters, it can be seen that the the following vectors lie in the tangent space to the GTR matrices at $Q^H$: 
\begin{align*}
V_5 &= \left(\mathbf{e}_1\mathbf{e}_2^T + \mathbf{e}_2\mathbf{e}_1^T\right) - \left(\mathbf{e}_3\mathbf{e}_4^T + \mathbf{e}_4\mathbf{e}_3^T\right) ,\\
V_6 &= \left(\mathbf{e}_1\mathbf{e}_3^T + \mathbf{e}_3\mathbf{e}_1^T\right) + \left(\mathbf{e}_2\mathbf{e}_4^T + \mathbf{e}_4\mathbf{e}_2^T\right) ,\\
V_7 &= \left(\mathbf{e}_1\mathbf{e}_3^T + \mathbf{e}_3\mathbf{e}_1^T\right) - \left(\mathbf{e}_2\mathbf{e}_4^T + \mathbf{e}_4\mathbf{e}_2^T\right) ,\\
V_8 &= \left(\mathbf{e}_1\mathbf{e}_4^T + \mathbf{e}_4\mathbf{e}_1^T\right) + \left(\mathbf{e}_2\mathbf{e}_3^T + \mathbf{e}_3\mathbf{e}_2^T\right) ,\\
V_9 &= \left(\mathbf{e}_1\mathbf{e}_4^T + \mathbf{e}_4\mathbf{e}_1^T\right) - \left(\mathbf{e}_2\mathbf{e}_3^T + \mathbf{e}_3\mathbf{e}_2^T\right) .
\end{align*}
The vectors $V_1,V_2,\ldots,V_9$ are linearly independent by construction. 
Standard linear algebra can be used to extend this to a basis $V_1,\ldots,V_{12}$ of $\mathbb{R}^{12}$.  

Next, the QR factorisation algorithm is applied to the $12\times 12$ matrix with columns $V_1,\ldots,V_{12}$ to obtain an orthonormal basis of tangent vectors $W_1,\ldots,W_{12}$ that is used to perturb $Q^H$. 
First, $Q^H$ is perturbed using $\nu_1,\ldots,\nu_5$ to obtain a GTR matrix $Q^R$ where, for $i \neq j$
\begin{equation*}
\log q_{ij}^R  = \log q_{ij}^H + \sum_{k=5}^9\nu_{k-4} W_{kij}, 
\end{equation*}
and the $\nu_k$ are independent $N(0,\sigma_R^2)$ and
$W_{kij}$ is the $(i,j)$-th element of the $4\times 4$ matrix corresponding to $W_k$.
The choice of basis $W_1,\ldots,W_{12}$ ensures that this perturbation is locally orthogonal to the sub-set of HKY85 matrices, and that the perturbation is otherwise isotropic within the sub-set of GTR matrices. 
The second stage perturbs $Q^R$ into the space of non-reversible rate matrices using $\eta_1,\eta_2,\eta_3$: for $i \neq j$
\begin{equation*}
\log q_{ij}  = \log q_{ij}^R + \sum_{k=10}^{12}\eta_{k-9} W_{kij}, \end{equation*}
and the $\eta_k$ are independent $N(0,\sigma_N^2)$ quantities.
This perturbation is locally perpendicular to the sub-set of GTR matrices in $M$. 
The equation determines the off-diagonal elements of the non-reversible rate matrix $Q$, while the diagonal elements are fixed in order to make the row sums zero. 
The size of the perturbation variance $\sigma_R^2$ can be thought of as representing the extent to which the rate matrix $Q$ departs from the class of HKY85 models remaining within the class of reversible models, while $\sigma_N^2$ represents the extent to which $Q$ departs from being reversible. \label{pg:2.3iii}

\section{Appendix 2}

The root on the majority rule consensus tree and the mode of the posterior distribution for root splits are different point summaries of the posterior distribution for root positions. Both can be approximated from posterior samples of rooted topologies but they need not coincide. For example, suppose the posterior output comprises the following five trees:

\begin{center}
\begin{tabular}{ l  l } \hline 
Tree 1:  &((A,B),(((E,F),D),C));\\
Tree 2:  &(((A,B),C),((E,F),D));\\
Tree 3:  &((((A,B),C),D),(E,F));\\
Tree 4:  &(((((A,B),C),D),E),F);\\
Tree 5:  &((A,B),(((E,F),D),C));\\ \hline
\end{tabular}
\end{center}


\begin{figure} [t!]	
\centering	
\includegraphics[scale=0.8]{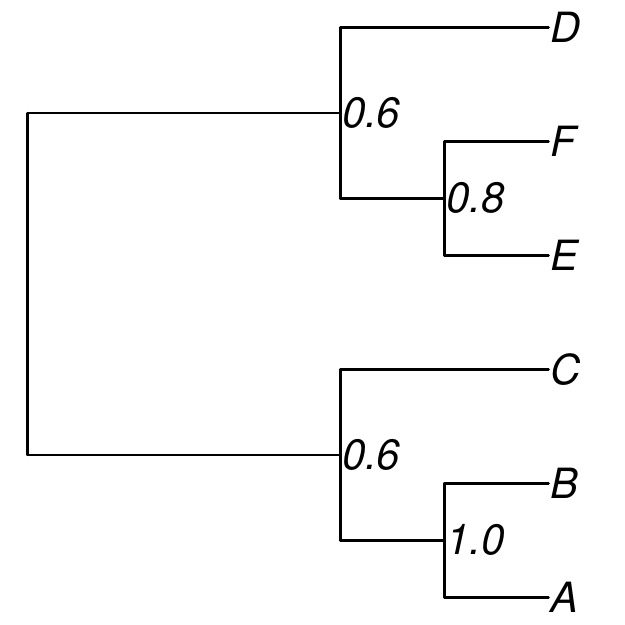}
\vspace{10.0pt}
\caption{Majority rule consensus tree for illustrative example.}\label{fig:app2_example}
\end{figure}

The clade (A,~B) appears on all the trees, and so is included in the consensus tree with probability one. Similarly, the clade (A,~B,~C)  appears on three trees (Tree 2, Tree 3 and Tree 4), and so appears in the consensus tree with support 0.6. Continuing in this fashion, the consensus tree is completed by incorporating the clades (E,~F) and (D,~E,~F) that appear with support 0.8 and 0.6 respectively. Hence, the root position on the consensus tree (displayed in Figure~\ref{fig:app2_example}) separates the taxa A,~B,~C from D,~E,~F.
On the other hand, the posterior for root splits is given in Table~\ref{tab:app2_example}. Clearly the posterior modal root split is (A,~B)~:~(C,~D,~E,~F)  which does not match the root split (A,~B,~C)~:~(D,~E,~F) on the consensus tree. 

\begin{table}[t!]
\caption{Posterior for root splits in illustrative example.}
\label{tab:app2_example}
\begin{center}
\begin{tabular}{ l c  c } \hline                                   
  \textbf{Root split} &\textbf{Count}  & \textbf{Probability}   \\ \hline
   (A,~B)~:~(C,~D,~E,~F)   & 2  & 0.4\\ 
   (A,~B,~C)~:~(D,~E,~F) & 1 & 0.2\\ 
   (E,~F)~:~(A,~B,~C,~D) & 1 & 0.2\\ 
   (F)~:~(A,~B,~C,~D,~E) & 1 & 0.2\\ \hline                  
\end{tabular}
\end{center}
\end{table}


%

\section{Acknowledgments}
This work was supported by the European Research Council (Advanced Investigator Award, grant number ERC-2010-AdG-268701, supporting S.C., S.E.H., T.A.W. and T.M.E.); and the Wellcome Trust (Program Grant, number 045404, to T.M.E.)

\bibliographystyle{plainnat}
\bibliography{references}




\end{document}